\documentclass[twocolumn]{aastex63}  
\usepackage{epsfig}
\usepackage{graphicx}
\usepackage{float} 
\usepackage{amsmath}
\usepackage{color}
\usepackage{amssymb}
\usepackage{amsfonts}
\usepackage{units}
\usepackage{bm}
\usepackage{longtable}
\usepackage{booktabs}
\usepackage[mathscr]{euscript}
\usepackage{soul}

\def\be{\begin{eqnarray}}
\def\ee{\end{eqnarray}}

\shorttitle{Multiwavelength Analysis of GRB 250101A}
\shortauthors{Du et al.}  

\begin{document}

\title{Multiwavelength Analysis of GRB 250101A: From Gamma-ray Prompt Emission to Optical Afterglow}

\correspondingauthor{Yuan-Pei Yang (ypyang@ynu.edu.cn), Jun Yang (jyang@smail.nju.edu.cn), Jinghua Zhang (zhang\_jh@ynu.edu.cn), Xiaowei Liu (x.liu@ynu.edu.cn)}

\author[0000-0002-8109-7152]{Guowang Du}
\affiliation{South-Western Institute for Astronomy Research, Yunnan University, Kunming, Yunnan 650504, People's Republic of China}

\author[0000-0001-8278-2955]{Yehao Cheng}
\affiliation{South-Western Institute for Astronomy Research, Yunnan University, Kunming, Yunnan 650504, People's Republic of China}

\author[0000-0001-6374-8313]{Yuan-Pei Yang}
\affiliation{South-Western Institute for Astronomy Research, Yunnan University, Kunming, Yunnan 650504, People's Republic of China}

\author[0000-0002-5485-5042]{Jun Yang}
\affiliation{School of Astronomy and Space Science, Nanjing University, Nanjing 210093, People's Republic of China}
\affiliation{Key Laboratory of Modern Astronomy and Astrophysics (Nanjing University), Ministry of Education, People's Republic of China}

\author[0000-0002-2510-6931]{Jinghua Zhang}
\affiliation{South-Western Institute for Astronomy Research, Yunnan University, Kunming, Yunnan 650504, People's Republic of China}

\author[0009-0008-5929-6658]{Dan Zhu}
\affiliation{South-Western Institute for Astronomy Research, Yunnan University, Kunming, Yunnan 650504, People's Republic of China}

\author[0009-0002-7625-2653]{Yu Pan}
\affiliation{South-Western Institute for Astronomy Research, Yunnan University, Kunming, Yunnan 650504, People's Republic of China}

\author[0009-0006-1010-1325]{Yuan Fang}
\affiliation{South-Western Institute for Astronomy Research, Yunnan University, Kunming, Yunnan 650504, People's Republic of China}

\author[0009-0006-5847-9271]{Xingzhu Zou}
\affiliation{South-Western Institute for Astronomy Research, Yunnan University, Kunming, Yunnan 650504, People's Republic of China}

\author[0000-0001-7225-2475]{Brajesh Kumar}
\affiliation{South-Western Institute for Astronomy Research, Yunnan University, Kunming, Yunnan 650504, People's Republic of China}

\author[0000-0001-5737-6445]{Helong Guo}
\affiliation{South-Western Institute for Astronomy Research, Yunnan University, Kunming, Yunnan 650504, People's Republic of China}

\author[0009-0003-6936-7548]{Xufeng Zhu}
\affiliation{South-Western Institute for Astronomy Research, Yunnan University, Kunming, Yunnan 650504, People's Republic of China}

\author[0009-0000-7791-8192]{Yangwei Zhang}
\affiliation{South-Western Institute for Astronomy Research, Yunnan University, Kunming, Yunnan 650504, People's Republic of China}

\author[0009-0009-9343-090X]{Fanchuan Kong}
\affiliation{South-Western Institute for Astronomy Research, Yunnan University, Kunming, Yunnan 650504, People's Republic of China}

\author[0009-0002-9069-8774]{Chenxi Shang}
\affiliation{South-Western Institute for Astronomy Research, Yunnan University, Kunming, Yunnan 650504, People's Republic of China}

\author[0009-0000-4068-1320]{Xinlei Chen}
\affiliation{South-Western Institute for Astronomy Research, Yunnan University, Kunming, Yunnan 650504, People's Republic of China}

\author[0000-0003-0394-1298]{Xiangkun Liu}
\affiliation{South-Western Institute for Astronomy Research, Yunnan University, Kunming, Yunnan 650504, People's Republic of China}

\author[0000-0003-1295-2909]{Xiaowei Liu}
\affiliation{South-Western Institute for Astronomy Research, Yunnan University, Kunming, Yunnan 650504, People's Republic of China} 

\begin{abstract}

The interaction between the relativistic jet and the circumburst medium produces a multiwavelength afterglow of a gamma-ray burst (GRBs). In this work, we present multiwavelength properties of GRB~250101A based on the observations of Swift, Fermi and Mephisto. The spectral analysis of Swift/BAT and Fermi/GBM reveals a soft prompt spectrum with a low-energy photon index of $-1.18$ and a peak energy of 33 keV, and the isotropic energy is $1.4\times10^{52}~{\rm erg}$. The prompt emission of GRB 250101A aligns with Type II GRBs in the Amati relation. Meanwhile, our analysis indicates that GRB 250101A is an X-ray-rich or X-ray-dominated GRB, with intrinsic properties suggesting that it is relatively softer than most classical GRBs. Optical observation with Mephisto, beginning 197 s post-trigger, shows a single power-law decay in $uvgriz$ bands, with $F_{\nu,\mathrm{obs}} \propto t^{-0.76} \nu^{-1.21}$. The observed spectral index significantly exceeds theoretical predictions under standard afterglow models, suggesting a color excess of $\sim0.216$ mag. However, combining X-ray and optical afterglow, we find that GRB 250101A is more likely a ``normal burst'' rather than an ``optical-dark burst'', and the dust extinction effect plays an important role in the optical blue bands. Furthermore, there is a structural change at $T_0+2924$ s in the optical light curve, indicating a density drop of $\sim50$ \% in the interstellar medium at a distance of $\sim0.13~{\rm pc}$. Our analysis shows that this GRB clearly shows some unique characteristics in its observed X-ray rich prompt emission as well as the circumburst environment, implying a special progenitor.

\end{abstract} 

\keywords{Gamma-ray bursts (629); Light curves (918); Optical astronomy (1776)}

\section{Introduction}\label{intro} 

Gamma-ray bursts (GRBs) are among the most energetic and luminous phenomena in the universe, believed to arise from relativistic jets powered by catastrophic events. Based on their durations, GRBs are generally classified as long GRBs and short GRBs \citep{Kouveliotou93}. Long GRBs are thought to result from the core collapse of rapidly rotating massive stars, with these events often linked to broad-line Type Ic supernovae \citep[e.g.,][]{Galama98, Woosley06}. In contrast, short GRBs are generally associated with the merger of compact objects, typically involving at least one neutron star, such as binary neutron star systems or binaries consisting of a neutron star and a black hole \citep[e.g.,][]{Eichler89, Paczynski91}.
The term ``afterglow'' refers to the emission phase following the prompt GRB, characterized by a decaying brightness. It arises from the interaction between the relativistic jets and the surrounding circumburst medium, which may include the interstellar medium (ISM) or the stellar wind from the progenitor. As the relativistic ejecta decelerated by the circumburst medium, two relativistic shock fronts will be generated: a forward shock that moves into the circumburst medium and a reverse shock that propagates back into the ejecta. These shocks accelerate particles in their respective regions, leading to the emission of multiwavelength nonthermal radiation through synchrotron processes \citep{Rees92, Meszaros93, Paczynski93, Katz94, Meszaros97, Sari98}.
Observations of GRB afterglows reveal complex behaviors \citep[e.g.,][]{Akerlof99, Harrison99, Berger03} that suggest the involvement of intricate physical conditions, such as non-uniform density distributions within the surrounding medium \citep{Dai98, Chevalier00}, ongoing energy injection into the blastwave \citep{Dai98b, Zhang01}, and the combined emission from both forward and reverse shocks \citep{Meszaros97, Kobayashi03, Zhang03}, among others.

The duration of the multiwavelength emission from GRB afterglows, spanning from gamma-ray to radio frequencies, varies from seconds to days. To study the properties of both the GRB central engine and the circumburst medium, it is essential to gather information in both temporal and spectral domains.
On January 1, 2025, the Swift Burst Alert Telescope (BAT) triggered on GRB~250101A at 13:22:50 UT \citep{38752, 38761, 38773, 38775, 38782}. The light curve from Swift/BAT revealed a complex structure, reaching a peak count rate of about 400 counts per second in the 15–350 keV range. The X-ray Telescope (XRT) initiated observations 127.4 seconds after the BAT trigger, with its location 151 arcseconds away from the BAT onboard position. The Ultra-Violet and Optical Telescope (UVOT) took a 150-second finding chart exposure with the white filter starting 132 seconds after the BAT trigger. An afterglow was detected at the coordinates ${\rm RA}=37.06873$ deg and ${\rm Dec}=+19.19442$ deg, with a 90\%-confidence error radius of about 1.1 arcseconds, which is 3.8 arcseconds from the center of the XRT error circle. 
Following the trigger, numerous observatories worldwide conducted follow-up observations of GRB~250101A, including GMG \citep{38753, 38776}, GIT \citep{38754}, Nanshan/HMT \citep{38755}, MASTER \citep{38756}, SVOM/C-GFT \citep{38758}, Mephisto \citep{38774}, Xinglong-2.16m \citep{38759}, SAO RAS \citep{38762, 38765, 38780}, AKO \citep{38763}, BOOTES-4/MET \citep{38764}, OASDG \citep{38768}, LCOGT \citep{38769, 38810}, Montarrenti Observatory \citep{38771}, NUTTelA-TAO/BSTI \citep{38777, 38790}, LCO \citep{38779}, Liverpool Telescope \citep{38798}, 1.3m DFOT \citep{38818}, REM \citep{38831}, SYSU 80cm telescope \citep{38837}, Terskol K-800 \citep{38869}, and CMO RC-500 \citep{38890}.
The redshift of GRB 250101A was determined to be $z = 2.481$, based on the detection of the Ly-$\alpha$ absorption feature at $\sim 4250$ \AA~and multiple metal absorption lines, including Si II at 1303 \AA, C II at 1335 \AA, Si IV at 1394 \AA, Si IV at 1403 \AA, C IV at 1549 \AA, Al II at 1671 \AA, Al III at 1855 \AA, and Al III at 1863 \AA \citep{38776}.
Notably, the early optical afterglow of GRB 250101A exhibited a rise in brightness at the early phase, peaking at $m_{g'} = 17.3$ mag and $m_{r'} = 16.6$ mag around 240 seconds post-burst, as observed by NUTTelA-TAO/BSTI \citep{38777}. 

In this study, we present and analyze the multiwavelength properties of GRB~250101A, based on observations made with Fermi/GBM, Swift/BAT, Swift/XRT, and the Multi-channel Photometric Survey Telescope (Mephisto) and the published NUTTelA data. We take the Swift/BAT trigger time of 13:22:50 UT on January 1, 2025, as the reference time $T_0$. All times in the results and figures are referenced to $T_0$.
The paper is organized as follows: Section \ref{observation} presents the measurements of the prompt gamma-ray emission (Section \ref{prompt}) and the multiwavelength afterglow (Section \ref{afterglow}) of GRB~250101A. Section \ref{physics} provides an analysis of the physical properties of the prompt emission (Section \ref{physics_prompt}) and the afterglow (Section \ref{physics_afterglow}). Finally, the results are discussed and summarized in Section \ref{conclusion}.
In all instances, we adopt the convention $Q_x = Q / 10^x$ in cgs units, unless otherwise noted.

\section{Multiwavelength observation of GRB~250101A}\label{observation}

\subsection{Observations of Prompt emission of GRB 250101A}\label{prompt}

\begin{figure}
    \centering
    \includegraphics[width=1.0\linewidth]{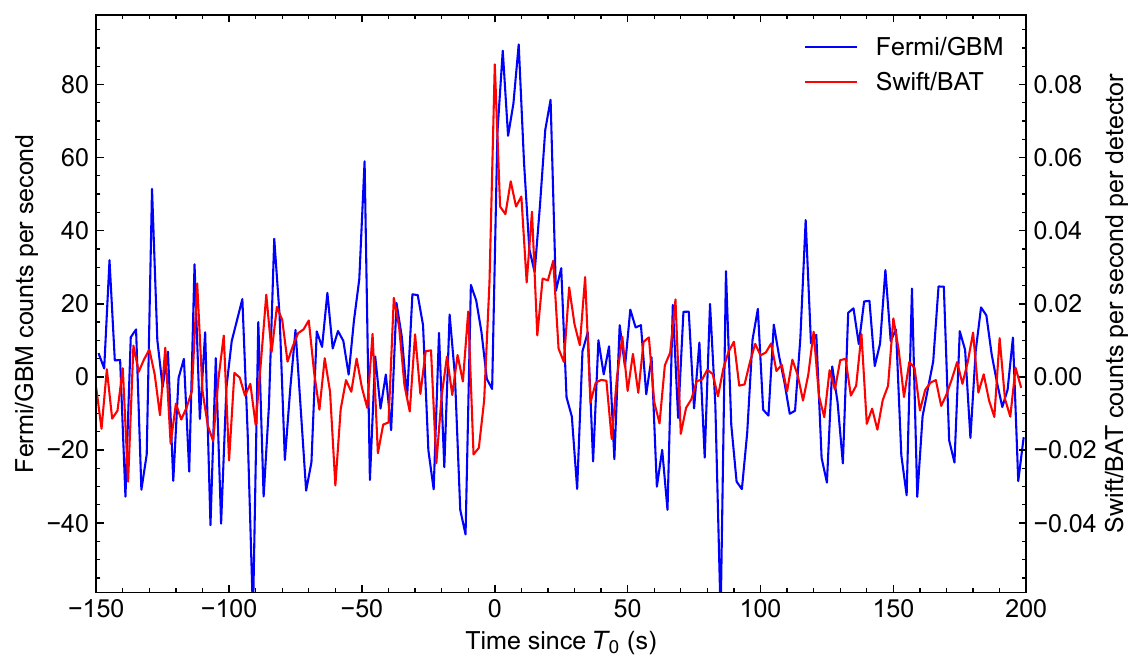}
    \caption{The light curves derived from Swift/BAT and Fermi/GBM data. The Fermi/GBM light curve is obtained from the data of detector n8 and has been background-subtracted. Both the light curves are binned with a time bin size of 2 s.}
    \label{fig:prompt_lc}
\end{figure}

GRB 250101A was initially triggered by Swift/BAT on January 1, 2025, at 13:22:50 UTC \citep{38752}. We downloaded the BAT data for GRB 250101A from the Swift Archive Download Portal\footnote{\url{https://www.swift.ac.uk/swift\_portal/}}. Following the standard BAT data analysis procedures\footnote{\url{https://www.swift.ac.uk/analysis/bat/index.php}} and utilizing \textit{HEASoft} tools (version 6.34), we performed the BAT data analysis. Specifically, the energy conversion of the BAT event data was corrected using the command \textit{bateconvert}. Subsequently, the commands \textit{batbinevt}, \textit{bathotpix}, and \textit{batmaskwtevt} were employed to produce the quality map, define hot pixels, and apply mask-weighting, respectively. Using \textit{batbinevt}, the BAT light curve was constructed for the time range of [-150, 200] s and the energy band of 15--350 keV, with a bin size of 2 s, as shown in Figure \ref{fig:prompt_lc}. The duration $T_{\rm 90, BAT}$ was determined from this light curve using the command \textit{battblocks}, yielding a value of 
\begin{align}
    T_{90,{\rm BAT}}=32\pm9~{\rm s}.
\end{align}

\begin{figure}
    \centering
    \includegraphics[width=1.0\linewidth]{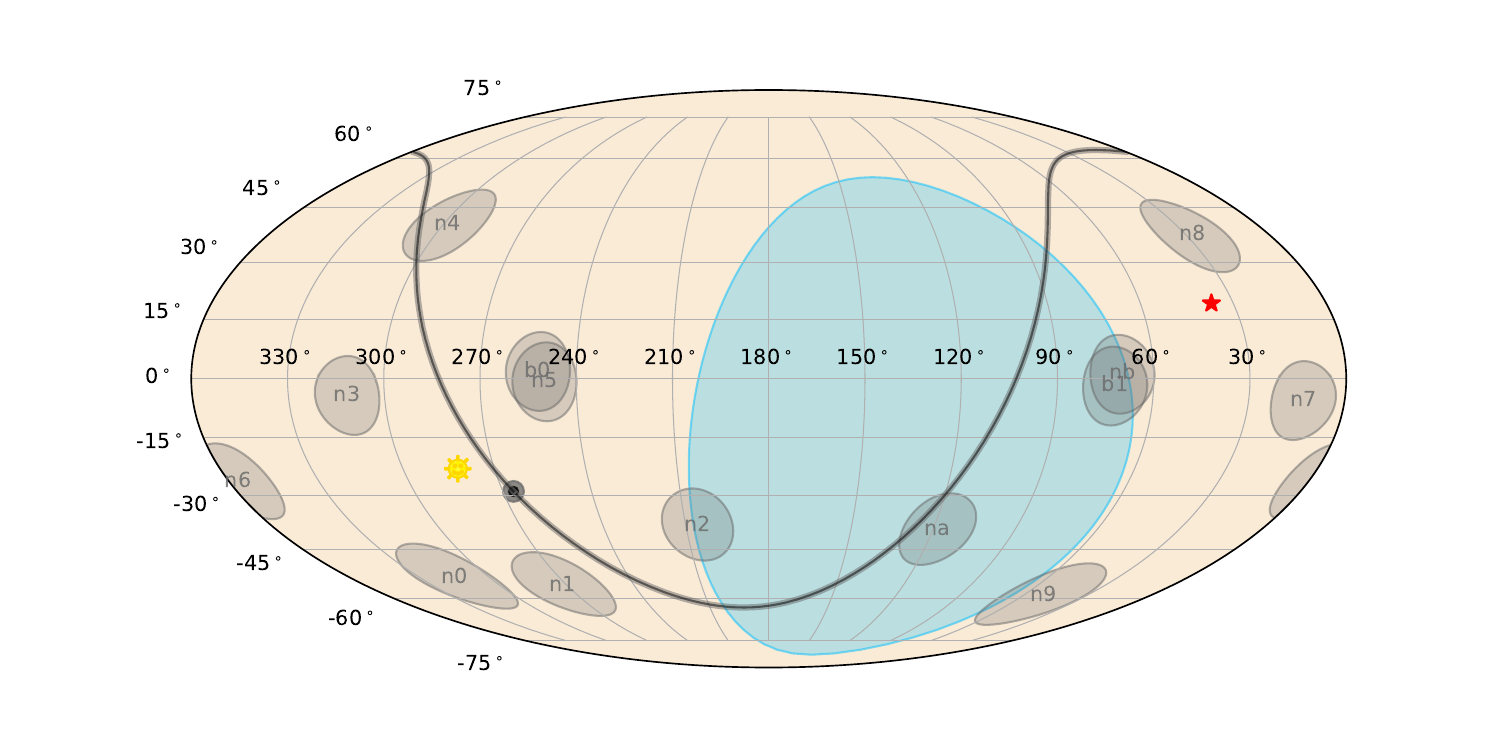}
    \caption{The Fermi/GBM sky map at $T_0$. The location of GRB 250101A is marked by a red star, while the yellow smiley face represents the Sun. The large blue region corresponds to the Earth as observed by Fermi. The gray-labeled circles denote the GBM detector pointings. The curved gray line across the plot indicates the Galactic plane, with a dot marking the Galactic center.}
    \label{fig:sky_map}
\end{figure}

Following the Swift/BAT trigger, we conducted a targeted search for the counterpart of GRB 250101A in historical archive data of Fermi/GBM\footnote{\url{https://heasarc.gsfc.nasa.gov/FTP/fermi/data/gbm/daily/}}. First, we downloaded the position history file from the Fermi/GBM data archive, which provides the spacecraft's orbital location and pointing information. This allows us to determine whether a source was visible at a given time (i.e., not blocked by the Earth) and, if visible, to identify the closest detector to the source. Figure \ref{fig:sky_map} presents the sky map at $T_0$, generated using GBM Data Tools \citep{GbmDataTools}, overlaid with the pointing directions of GBM detectors, the Swift/XRT localization of GRB 250101A, and the Earth as observed by Fermi. It is evident that GRB 250101A was visible to GBM (i.e., not in the Earth's shadow) and was located far from the Sun’s position, ensuring that their gamma-ray emissions would not be confused or indistinguishable. We note that detectors n8, n7, nb, and b1 maintained a minimal angular separation of less than 60° from the burst, with specific angles of 20°, 34°, 37°, and 40°, respectively. 

Subsequently, we downloaded the time-tagged event (TTE) dataset covering the time of GRB 250101A from the Fermi/GBM data archive. Data analysis of TTE was performed following the procedures outlined in \citet{Yang2020ApJ, Yang2022Natur}. Interestingly, in detectors n8, n7, and nb, we identified a faint, untriggered transient source whose timing and profile closely matched the BAT light curve of GRB 250101A, as shown in Figure \ref{fig:prompt_lc}. We applied the Bayesian Blocks algorithm \citep{Scargle2013ApJ} to the light curve of detector n8 within the time interval of [-150, 200] s, using a time bin size of 2 s. This analysis identified a block with the highest signal-to-noise ratio within the time range of [0, 20] s. Following the methodology of \citet{Vianello2018ApJS}, the signal significance of this block was calculated to be approximately 9$\sigma$. Based on this high significance, we conclude that this transient source detected in the Fermi/GBM data is genuine and represents the counterpart of GRB 250101A in Fermi/GBM observations. The burst duration $T_{\rm 90, GBM}$, determined by the time width between the 5\% and 95\% levels of the total accumulated net photons of detector n8 in the energy range of 10--1000 keV, is calculated to be 
\begin{align}
    T_{90,{\rm GBM}}=20\pm1~{\rm s}.
\end{align}

We performed a time-integrated joint spectral analysis of GRB 250101A using Swift/BAT and Fermi/GBM data within the time interval [-1, 31] s. For the Swift/BAT data, we utilized the \textit{HEASoft} commands as follows: \textit{batbinevt} to generate the BAT spectrum, \textit{batphasyserr} and \textit{batupdatephakw} to apply systematic errors and correct the ray-tracing keywords, and \textit{batdrmgen} to generate the BAT response matrix file. For the Fermi/GBM data, we extracted the total spectrum, background spectrum, and detector response matrix for detectors n8, n7, nb, and b1. The total and background spectra were obtained by summing the total photon counts and background photon counts in each energy channel, respectively. The response matrices were generated using the Python module \textit{gbm\_drm\_gen} \citep{Burgess_2018MNRAS, Berlato_2019ApJ}. We then performed spectral fitting using \textit{MySpecFit} \citep{Yang2022Natur, Yang2023ApJ}, which employs \textit{MultiNest} \citep{Feroz_2008MNRAS, Feroz_2009MNRAS, Buchner_2014A&A, Feroz_2019OJAp} as a Bayesian sampler for Bayesian inference. For the BAT spectrum, we used the $\chi^2$ statistic to evaluate the likelihood between the data and the model, while for the GBM spectrum, we used the PGSTAT \citep{Arnaud_1996ASPC} statistic. The cutoff power law (CPL) model was adopted to fit the time-integrated spectrum, which can be expressed as 
\begin{align}
    N(E)=A_{\gamma}E^{\alpha_{\gamma}}{\rm exp}(-E/E_{\rm c}),
\end{align}
where $\alpha_{\gamma}$ is the low-energy photon spectral index of the prompt gamma-ray emission, and the relationship between the peak energy $E_{\rm p}$ and the cutoff energy $E_{\rm c}$ is given by $E_{\rm p}=(2+\alpha_{\gamma})E_{\rm c}$. The best-fit parameters from the joint spectral fitting, along with their 1$\sigma$ uncertainties are
\begin{align}
    \alpha_{\rm \gamma} = -1.18_{-0.43}^{+0.30}~~~ {\rm and}~~~E_{\rm p} = 33_{-6}^{+5}~{\rm keV}.
\end{align}
Based on the spectral fitting results, we calculated the average flux of GRB 250101A to be $2.9_{-0.4}^{+0.9}\times10^{-8}~{\rm erg~cm^{-2}~s^{-1}}$. Given its redshift of $z = 2.481$ \citep{38776}, the isotropic-equivalent energy of GRB 250101A was determined to be 
\begin{align}
    E_{\rm \gamma,iso}=1.4_{-0.2}^{+0.4}\times10^{52}~{\rm erg}.\label{Egamma}
\end{align}
The detailed physical properties of the prompt gamma-ray emission will be discussed in Section \ref{physics_prompt}.

\subsection{Observations of Optical afterglow of GRB~250101A}\label{afterglow}
The photometric observations of GRB~250101A were conducted using the Mephisto telescope, a pioneering instrument managed by the South-Western Institute for Astronomy Research at Yunnan University. Located at the Lijiang Observatory (IAU code: 044) of the Yunnan Astronomical Observatories, Chinese Academy of Sciences, Mephisto is recognized as the first wide-field, multi-channel photometric survey telescope. Currently in its commissioning phase, Mephisto is equipped with two Andor Technology single-chip CCD cameras for the blue ($uv$) and yellow ($gr$) channels, respectively. The spatial sampling of the images in these channels is $0^{\prime\prime}.429/\text{pix}$. The red channel ($iz$) utilizes a single-chip camera developed at the National Astronomical Observatories of China (NAOC). The spatial sampling of the red channel images is $0^{\prime\prime}.286/\text{pix}$. Each channel has a field of view of approximately 43 arcmin × 43 arcmin. Plans are underway to expand the field of view to 2 square degrees with the implementation of mosaic cameras, expected in the spring of 2025. The three CCD cameras enable simultaneous imaging in the $ugi$ or $vrz$ bands for each observation. The wavelength ranges of the $u$, $v$, $g$, $r$, $i$, and $z$ filters are 320–365, 365–405, 480–580, 580–680, 775–900, and 900–1050 nm, respectively, with central wavelengths at 345, 385, 529, 628, 835, and 944 nm, respectively \citep{Yang24, Chen2024ApJ}.
A comparison of the transmission curves of the Mephisto filters and those of other systems is shown in \citet{Yang24}.  

The simultaneous multiband photometric observations with Mephisto were initiated 197 seconds after the Swift/BAT trigger. The exposure times of the three channels were adjusted to compensate for the different bandwidths and efficiencies of the three channels for dense sampling and better signal-to-noise ratios (SNRs). From 13:26:07 to 14:13:55 UTC on 2025-01-01, initial exposure times were 180, 50, and 79 seconds in the blue ($uv$), yellow ($gr$), and red ($iz$) channels, respectively. During that time, 7, 5, 22, 15, 14, and 10 frames were collected in the $u,v,g,r,i$, and $z$ bands. Subsequently, the exposure times were increased to 300 seconds in the $u,v,g,r$ bands and to 294 seconds in the $iz$ bands from 14:15:41 to 17:44:03 UTC. In that time, 18, 16, 18, 16, 18, and 16 frames were collected in the $u,v,g,r,i$, and $z$ bands, respectively. In total, 25, 40, 32, 21, 31, and 26 frames were obtained in the $u,v,g,r,i$, and $z$ bands, respectively, during the first night.
From 11:33:33 to 12:12:07 UTC on 2025-01-02, exposures were taken in the $u$, $v$, $g$, and $r$ bands for 300 seconds each, and in the $i$ and $z$ bands for 294 seconds each. During this period, 3 frames were collected in each of the $u$, $v$, $g$, $r$, $i$, and $z$ bands. Additionally, on 2025-01-22, 3, 4, 3, 4, 3, and 4 frames were obtained in the $u$, $v$, $g$, $r$, $i$, and $z$ bands, respectively, with exposure times of 300 seconds per band, to capture images of the host galaxy. Note that the seeing conditions were poor on the last night, leading to large FWHM values, indicative of degraded resolution and the upper limit magnitudes for $uvgriz$ bands are 21.15, 22.01, 21.96, 22.38, 21.12 and 20.73 mag, respectively.

The raw frames were preprocessed using a specialized pipeline developed for Mephisto, which included bias and dark subtraction, flat-fielding, and cosmic ray removal. A minimum of 10 bias frames were acquired prior to the observing night and stacked using a pixel-wise trimmed average to generate the master bias image. A series of dark frames with varying exposure times were also acquired before the observing night. The primary dark image for each exposure time was generated by applying a pixel-wise trimmed median, using at least 3 dark frames. Subsequently, a minimum of 10 qualified twilight flatfield frames for each filter were acquired before the observing night. These frames were normalized to a spatially trimmed global mean of one and then stacked using a pixel-wise trimmed average to create the master flat image. Afterward, a raw science image with an exposure time of less than 10 seconds was subtracted by the master bias image, while images with longer exposure times were subtracted by the primary dark image with corresponding exposure times. The result was then divided by the most recently generated flatfield calibration images for the corresponding filter. Cosmic rays were detected in the flatfield-calibrated image using \texttt{Astro-SCRAPPY}  \citep{curtis_mccully_2018}, an Astropy-affiliated package for rapid cosmic ray rejection based on a Laplacian edge detection method \citep{Dokkum2001PASP}. They were removed with Kernel-based interpolation.

The source detection was performed using \texttt{SExtractor} \citep{Bertin1996}. Astrometric calibration was then performed using reference stars from the Gaia DR3 catalog \citep{GaiaDR3}. A first approximation for astrometric solutions was obtained using the \texttt{Astrometry.net} module with no first guess \citep{Lang2010AJ}. It was subsequently refined with  \texttt{SCAMP} software \citep{Bertin2006ASPC}. The delivered uncertainties in the astrometric calibration were estimated to be better than 0.1 pixels.  

Aperture photometry was carried out using \texttt{SExtractor} \citep{Bertin1996}. For photometric calibration, we employed recalibrated Gaia BP/RP (XP) spectra through the Synthetic Photometry Method \citep{Huang2024ApJS, Xiao2023ApJS}. To calculate the synthetic magnitudes in the AB system, we convolved the XP spectra with the transmission curves corresponding to the Mephisto telescope across different bands. The XP spectra span wavelengths from 336 to 1020 nm, but this range does not fully cover the $u$ and $z$ bands of Mephisto. To resolve this, we performed a linear extrapolation of the XP spectra to fill in the missing wavelength regions. The resulting uncertainties for the synthetic $u$ and $z$ magnitudes due to this extrapolation were estimated to be 0.016 mag and 0.0007 mag, respectively. Then, we computed the magnitude difference, $\Delta m$, between the instrumental and synthetic magnitudes for high-quality, non-variable stars in each frame. To correct for any potential gain measurement inaccuracies, we calculated the weighted average of $\Delta m$ for each CCD output, applying this value for the photometric calibration of the targets located within that output. The weights for the averaging process were derived from the uncertainties including both the synthetic and instrumental magnitude error contributions. The uncertainties associated with the photometric calibration were determined to be better than 0.03 mag in the $u$ band, 0.01 mag in the $v$ band, and 0.005 mag in the $griz$ bands. Additionally, the transmission efficiency curves for each Mephisto band were derived based on the measured efficiency curves of individual optical path components, from the primary mirror to the detector.

\begin{figure}
\centering
\includegraphics[width = 1\linewidth, trim = 0 0 0 0, clip]{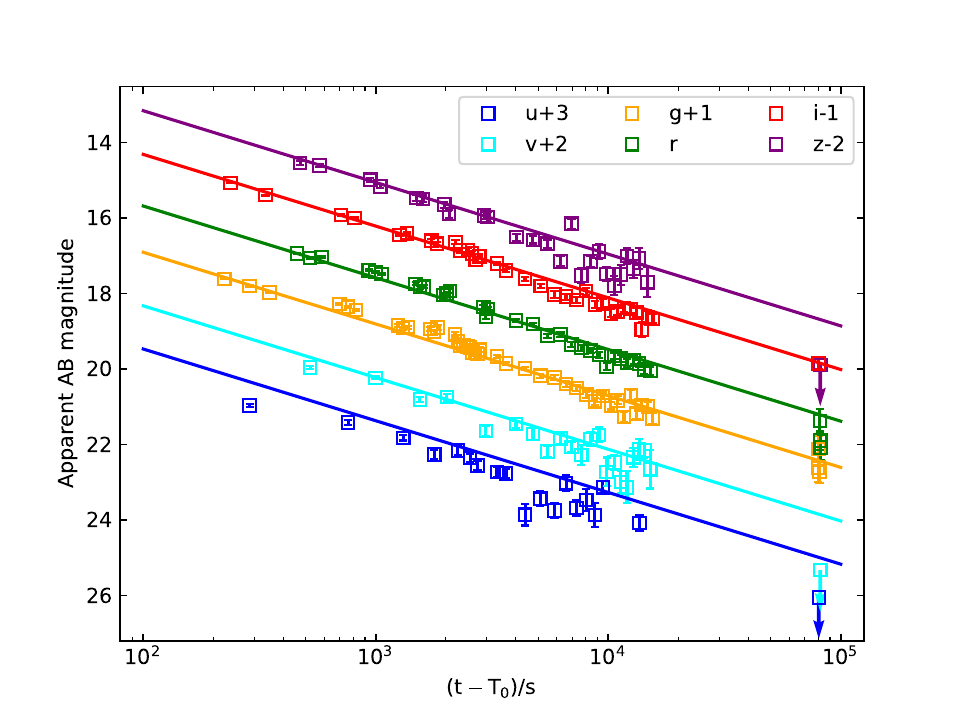}
\includegraphics[width = 1\linewidth, trim = 0 0 0 0, clip]{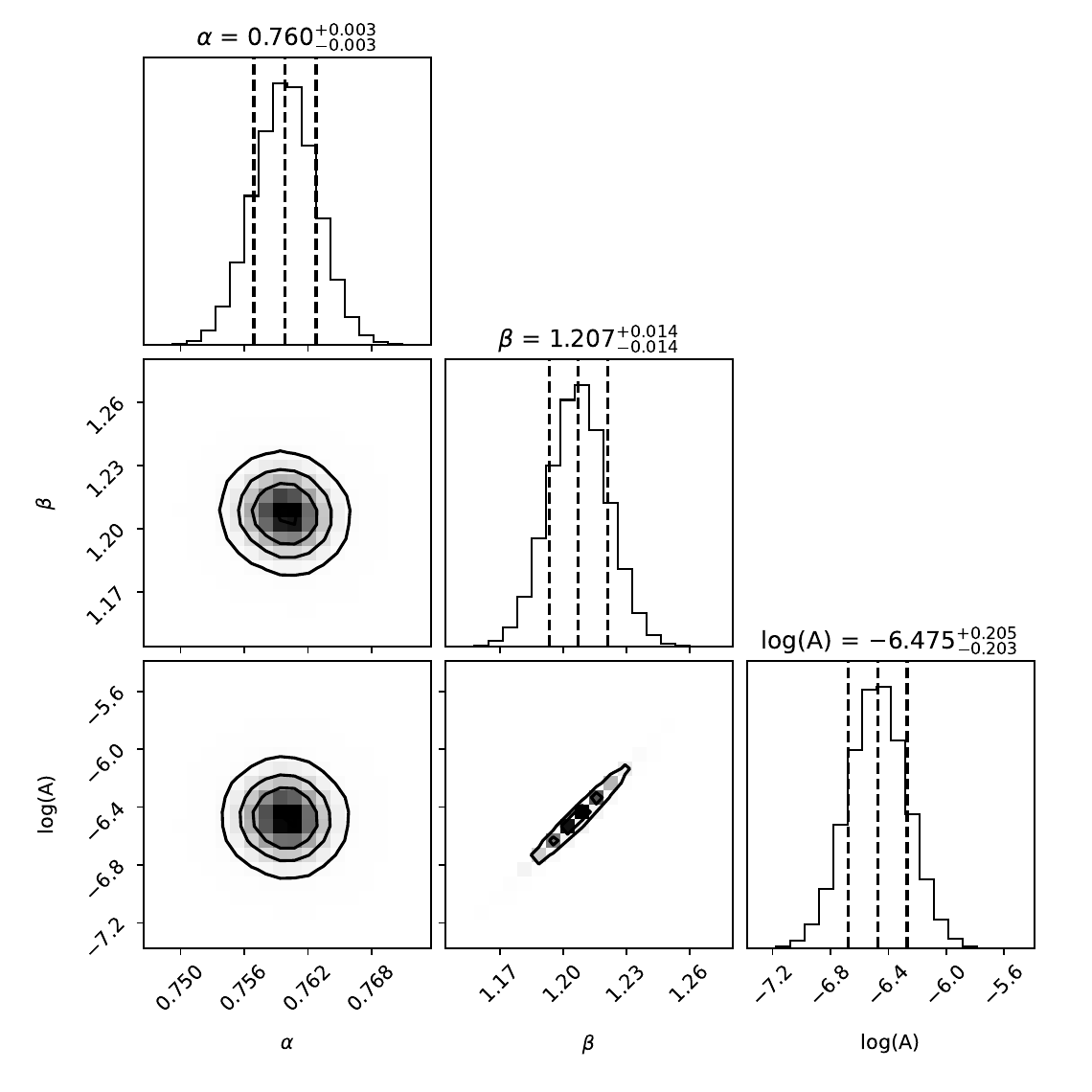}
\caption{Multiband light curves of the early optical afterglow of GRB~250101A obtained with Mephisto. The Galactic extinction has been corrected in this figure. Top panel: Best fit to the Mephisto multiband light curves of the optical afterglow of GRB~250101A in the $uvgriz$ bands. The solid lines of different colors denote the best-fitting curves given by Eq.(\ref{lc_spec}). 
The times correspond to the middle times of the exposures.
Bottom panel: Two-dimensional projections of the posterior probability distributions of the fitting parameters of the model given by Eq.(\ref{lc_spec}). Contours show the 0.5, 1, 1.5, and 2$\sigma$ significance levels.}\label{result} 
\end{figure}

For the first night observation, the target was marginally detected in the last few frames in the $u$ band and $z$ band. For the second night observation, the source was marginally detected in the $i$ band, while it was not detected in the $uvz$ bands. To improve the signal-to-noise ratio (SNR) for better detection, we co-added multiple exposures of those frames in the two bands using the \texttt{SWarp} program \citep{Bertin2010ascl}. The image with the smallest median FWHM was chosen as the reference for aligning the flux. The fluxes of the other images were adjusted to match the reference image by using stars with SNRs $>$20. The same procedures as previously outlined were applied to perform aperture photometry on the co-added images. Photometric calibration was performed using the zero-point values derived from the reference image. For sources that were not clearly detected, the 3-sigma limiting magnitudes were determined from the co-added images.
In Table \ref{1.6mdata_table} in the appendix, we show the photometry data of the optical afterglow of GRB~250101A in the $uvgriz$ band of Mephisto, and all measurements were in AB magnitudes and were not corrected for the Galactic extinction. The times in the table correspond to the middle time of each exposure. The uncertainties of the photometric calibration are estimated to be better than 0.03, 0.01, and 0.005 mag for $u$, $v$, and $griz$, respectively.

In the line of sight of GRB~250101A, the Galactic extinction is $E_{B-V}=0.141$ mag \citep{Schlegel98,Schlafly2011}. According to the Galactic extinction law from \cite{Fitzpatrick1999PASP}, $R_{V}=3.1$, the extinction values for $uvgriz$ bands of Mephisto are 0.697, 0.636, 0.454, 0.357, 0.227, 0.184 mag, respectively. 
In the top panel of Figure \ref{result}, we correct the Galactic extinction and present the observed light curves of the optical afterglow of GRB 250101A obtained with Mephisto in the $uvgriz$ bands.

\section{Physical Properties of the multiwavelength observations of GRB~250101A}\label{physics}

In this section, we will discuss the physical properties of the multiwavelength observations of GRB~250101A. The physical implications of the prompt emission and the afterglow will be discussed in Section \ref{physics_prompt} and Section \ref{physics_afterglow}, respectively.

\subsection{Prompt gamma-ray emission of GRB~250101A}\label{physics_prompt}

\begin{figure}
    \centering
    \includegraphics[width=1.0\linewidth]{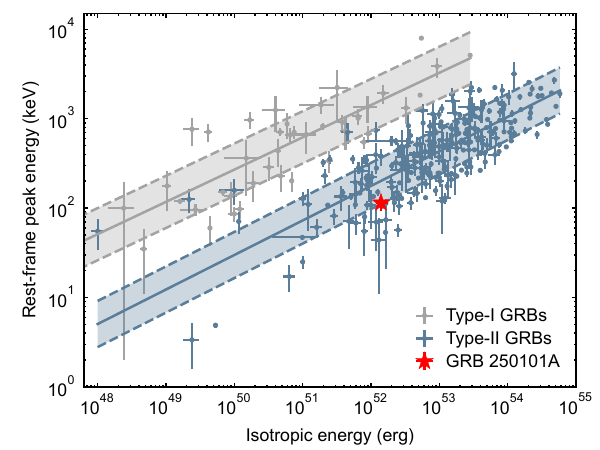}
    \caption{The $E_{\rm p}(1+z)$ -- $E_{\gamma,{\rm iso}}$ diagram. GRB 250101A is marked with a red star. The gray and blue data points represent the Type I (associated with compact star mergers) and Type II (originating from the core collapse of massive stars) GRB samples, respectively. The solid line indicates the best-fit linear model, while the shaded area represents its 1$\sigma$ intrinsic scatter region.}
    \label{fig:epz_eiso}
\end{figure}

In the $E_{\rm p}(1+z)$ -- $E_{\gamma,{\rm iso}}$ diagram (i.e., the Amati relation), GRBs from different physical origins generally follow distinct linear tracks: compared to Type I GRBs (associated with compact star mergers), Type II GRBs (originating from the core collapse of massive stars) typically exhibit lower rest-frame peak energies and higher isotropic-equivalent energies. 
In Figure \ref{fig:epz_eiso}, we present the $E_{\rm p}(1+z)$ -- $E_{\gamma,{\rm iso}}$ diagram using a sample of GRBs with known redshifts \citep{Amati2002A&A, Zhang2009ApJ, Zhang2018NatAs, Minaev2020MNRAS}. 
It is found that GRB 250101A lies well within the 1$\sigma$ region of the Type II GRB track, although its rest-frame peak energy is relatively lower than that of other Type II GRBs.

\begin{figure}
    \centering
    \includegraphics[width=1.0\linewidth]{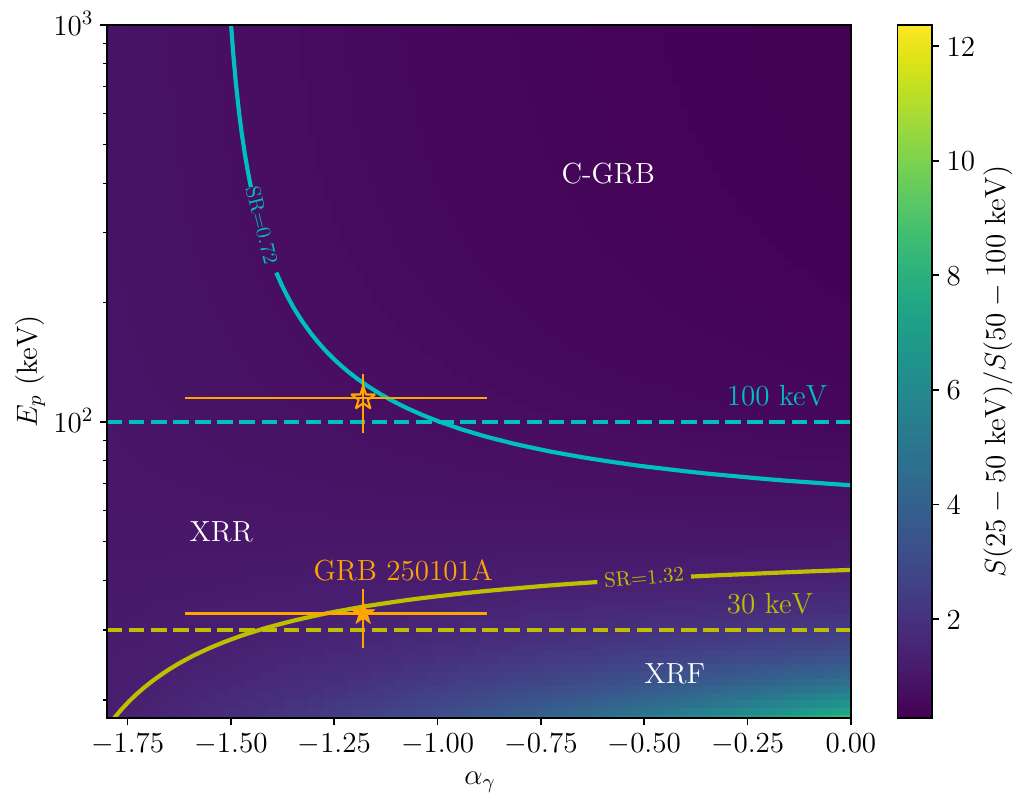}
    \caption{The fluence ratio distribution in the parameter space of the CPL model. The cyan and yellow solid lines represent SR = 0.72 and SR = 1.32, respectively, which are the dividing lines under the fluence ratio classification criterion. The cyan and yellow dashed lines correspond to 100 keV and 30 keV, respectively, which are the boundaries in the peak energy classification scheme. GRB 250101A is highlighted with a filled orange star, while the unfilled orange star represents GRB 250101A in the rest frame.}
    \label{fig:flnc_ratio} 
\end{figure}

Compared to other GRBs, GRB 250101A exhibits a much softer spectrum, which may suggest that it is not a classical GRB (C-GRB, where gamma-ray radiation dominates), but rather a GRB dominated by X-ray emission. Such subclasses are commonly known as X-ray Flashes (XRFs) or X-ray Rich GRBs (XRRs) \citep{Heise2003AIPC, Sakamoto2005ApJ}. XRRs typically display softer spectra and lower luminosities compared to C-GRBs, while XRFs represent even softer events with strong X-ray emission features, where the peak energy is generally less than 30 keV. Both types may be natural extensions of C-GRBs in the less-luminous, softer-spectral regime. However, the exact physical processes that lead to these events remain contentious.

To distinguish C-GRBs, XRRs, and XRFs based on their spectral and fluence properties, two classification criteria have been adopted: peak energy and fluence ratio (SR) between two different energy bands \citep{Sakamoto2005ApJ, Sakamoto2008ApJ}. In the peak energy classification scheme, 30 keV and 100 keV are commonly used as boundaries to separate these three categories. For the fluence ratio criterion, taking the 25--50 keV and 50--100 keV bands as an example, the classification is defined with SR = 0.72 and SR = 1.32 as the dividing lines. To incorporate both classification standards, we follow the approach of \citet{Yin2024ApJ} by calculating the distribution of fluence ratios in the CPL model parameter space ($\alpha_{\rm \gamma}$ vs. $E_{\rm p}$), as shown in Figure \ref{fig:flnc_ratio}. It is found that GRB 250101A lies almost exactly on the boundary between XRR and XRF, regardless of whether the classification is based on peak energy or fluence ratio criteria. Given that the redshift of this GRB is known, we can determine its intrinsic peak energy in the rest frame. Interestingly, in Figure \ref{fig:flnc_ratio}, GRB 250101A in the rest frame is positioned almost exactly on the boundary between XRR and C-GRB. Overall, the observational characteristics of GRB 250101A suggest that it is an X-ray-rich or X-ray-dominated GRB, with intrinsic properties indicating that it is relatively softer than most C-GRBs.

\subsection{Optical afterglow of GRB~250101A}\label{physics_afterglow}

A prevalent and significant characteristic of GRB afterglows is their ``multi-wavelength'' nature. The external shock model with synchrotron or synchrotron self-Compton radiation mechanisms predicts that an afterglow typically spans a very wide frequency range, from the low-frequency radio band all the way up to the gamma-ray band.  The broadband Spectral Energy Distribution (SED) of an afterglow at any given moment is proposed to follow a multi-segment broken power law. 
When observed at a specific frequency, the light curve of a GRB afterglow should also exhibit characteristics of a (broken) power law.

\subsubsection{Preliminary analysis in optical band}

As the blast wave decelerates in the surrounding medium, the afterglow light curves at all wavelengths are expected to decay in a power-law fashion after an initial rising phase. Consequently, the afterglow flux density can be characterized in the following manner:
\begin{align}
F_{\nu,{\rm obs}}(\nu, t)= A t^{-\alpha}\nu^{-\beta}+ F_{\nu,{\rm host}}(\nu),\label{lc_spec}
\end{align}
where $A$ is a constant, $\alpha$ the temporal decay index, $\beta$ the spectral index, $\nu$ the frequency, $F_{\nu,{\rm host}}(\nu)$ the SED of the host galaxy. 
The relation between the AB magnitude and the flux density is $m_\nu=-2.5\log F_{\nu,{\rm obs}}-48.6$.

The light curves of some optical afterglows of GRBs in their early phase (i.e., the first few hours) display a smooth, hump-like feature in its early light curve, and subsequently transitions into a typical power-law decay at later times with a decay index of $\sim-1$ \citep{Liang13}. This behavior aligns with predictions from the external forward shock model, where the early hump is interpreted as the onset of the afterglow, marking the deceleration radius of the blastwave. 

For the optical afterglow measured by Mephisto, based on Eq.(\ref{lc_spec}), we first apply the Bayesian inference to the measurements of GRB~250101A and derive the temporal decay index and the spectral index,
\begin{align}
\alpha_{\rm obs}=0.760\pm0.003~~~{\rm and}~~~\beta_{\rm obs}=1.207\pm0.014,
\end{align}
where the host flux is taken as $F_{\nu,{\rm host}}(\nu)\simeq0$ due to the non-detection of the host galaxy. Notably, the early optical afterglow of GRB 250101A exhibited a rise in brightness at the early phase, peaking at $m_{g'} = 17.3$ mag and $m_{r'} = 16.6$ mag around 240 seconds post-burst, as observed by NUTTelA-TAO/BSTI \citep{38777}. Although the simultaneous multiband photometric observations with Mephisto were initiated 197 seconds after the Swift/BAT trigger, we did not identify the rising phase due to taking a relatively long exposure time of 300 seconds.

\begin{figure}
\centering
\includegraphics[width = 1\linewidth, trim = 0 0 0 0, clip]{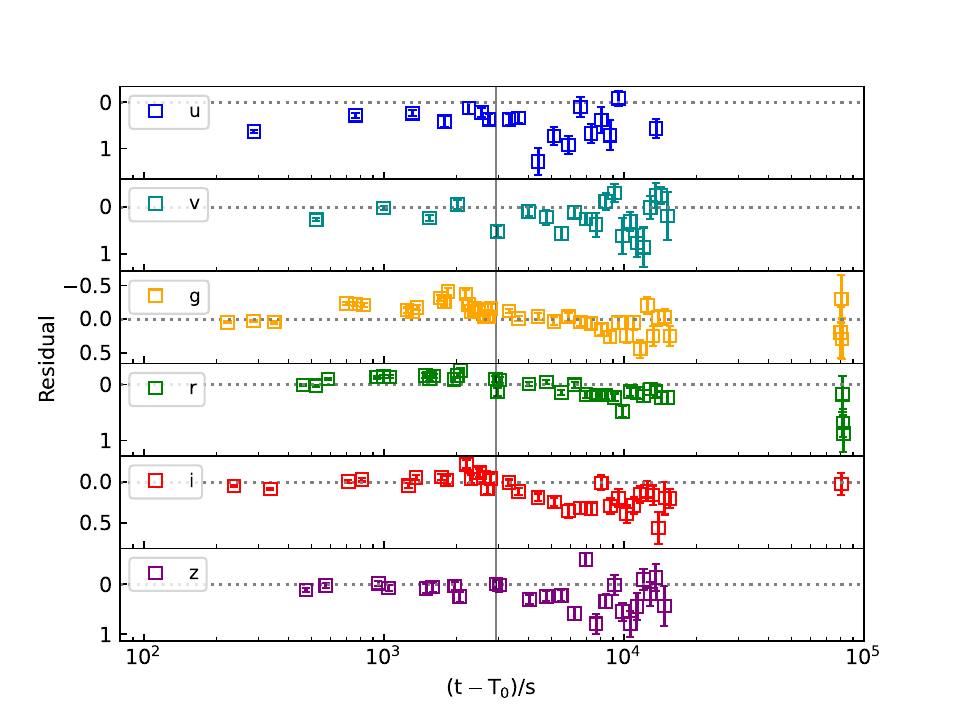}
\caption{The residual of Mephisto's measurement compared with the single power-law model shown in Figure \ref{result}. The vertical grey solid line represents the breakpoint time of $T_0+2924^{+112}_{-190}$ s, see Section \ref{break} for a detailed discussion.}\label{residual} 
\end{figure}

We notice that there are some biases between the measurements and the prediction of the standard afterglow model at the blue band of the Mephisto, as shown in the top panel of Figure \ref{result} and Figure \ref{residual}, suggesting a possible extinction from the host galaxy or even the circumburst medium. Such a feature has also been recently observed in GRB~240825A \citep{Cheng25}, which is attributed to the simultaneous photometry of the early afterglow in wide $uvgriz$ bands of Mephisto.
At the optical band, the main possible scenario for the standard afterglow model with uniform ISM and slow cooling has \citep{Cheng25,Gao13,Zhang18}
\begin{align}
\alpha=\frac{3(p-1)}{4},~~~\beta=\frac{p-1}{2}~~~{\rm and}~~~\beta(\alpha)=\frac{2\alpha}{3},\label{ab1}
\end{align}
where $\beta=2\alpha/3$ is the corresponding closure relation in this case.
Considering that the extinction from the circumburst medium might be significant, the measured temporal decay index $\alpha_{\rm obs}\simeq0.760$ is probably more robust than the measured spectral index $\beta_{\rm obs}\simeq1.207$. We obtain the distribution index of electrons, $p\simeq2.01$, and the resultant value of the spectral index (that is directly inferred from $\alpha_{\rm obs}$) is $\beta(\alpha_{\rm obs})\simeq0.51$.
We notice that our measured value of $\beta_{\rm obs}=1.207$ of GRB 250101A is significantly larger than the values of $\beta(\alpha_{\rm obs})\simeq0.51$,
suggesting that there is a significant extinction contribution from the circumburst medium. 

\begin{figure}
\centering
\includegraphics[width = 1\linewidth, trim = 0 0 0 0, clip]{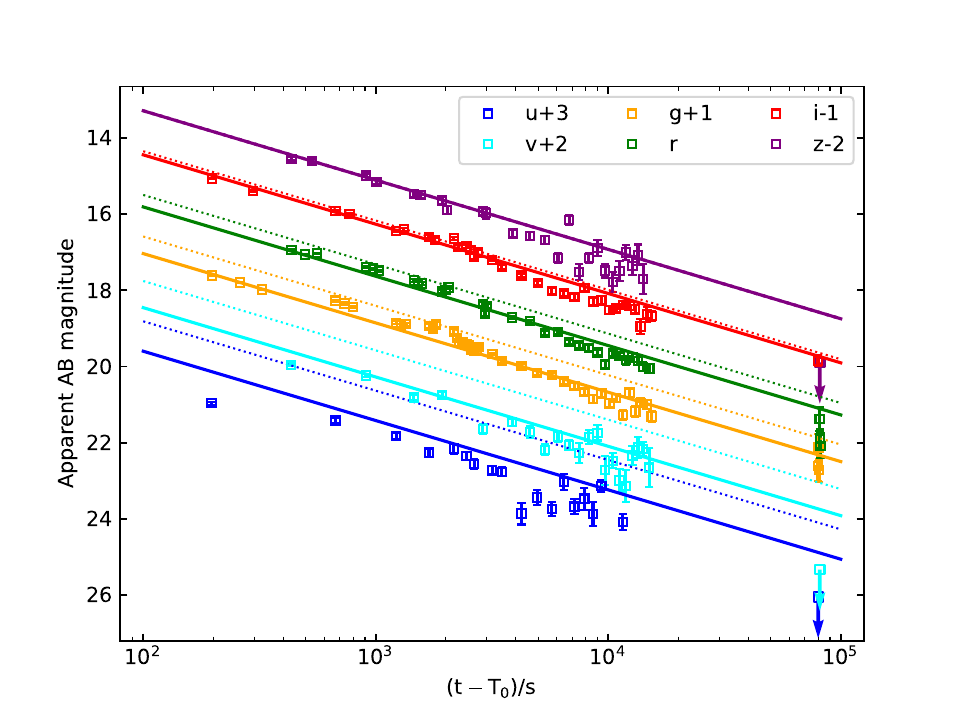}
\caption{Best-fitting light curves of the early optical afterglow of GRB~250101A with the color excess as a free parameter. The closure relation of $\beta=2\alpha/3$ is used. The best-fitting color excess is $E_{B-V}=0.216$ mag, and the best fitting temporal index is $\alpha=0.728$. The solid and dotted lines correspond to the model after and before the extinction correction, respectively. For the dotted line, we assume that the extinction at $z$ band could be neglect.}\label{lc_ext} 
\end{figure}

Following \citet{Cheng25}, we define the observed $j$-band magnitude as $m_j$ and the intrinsic magnitude that would be observed in the absence of dust as $m_{0,j}$, then the color excess in $(u-z)$ is
\begin{align}
E_{u-z}
=\Delta m_{uz}-\Delta m_{0,uz}\simeq1.09\left[\beta_{\rm obs}-\beta(\alpha_{\rm obs})\right].\label{Euz}
\end{align}
where $\beta(\alpha_{\rm obs})$ is given by Eq.(\ref{ab1}).
We assume that the host has an extinction law similar to that of the Milky Way\footnote{
While the extinction profiles of GRB afterglows are commonly modeled after the Small Magellanic Cloud (SMC) (as proposed in studies \citep[e.g.,][]{Zafar18}, within the optical wavelength range, the distinctions between the Milky Way and SMC extinction curves become relatively minor as demonstrated by earlier works \citep[see][]{Cardelli89,Gordon03}.
}. For the $u$ and $z$ bands, we have $R_u\simeq4.95$ and $R_z\simeq1.31$, according to the Galactic extinction law of \citet{Fitzpatrick1999PASP} with $R_{V}=3.1$. Thus, the color excess in $B-V$ could be estimated as,
\begin{align}
E_{B-V}=\frac{E_{u-z}}{R_u-R_z}.\label{EBV}
\end{align}
Using Eq.(\ref{lc_spec}), Eq.(\ref{Euz}) and Eq.(\ref{EBV}), we can refit the multiband light curves of the optical afterglow of GRB 250101A and obtain $\alpha$ and $E_{B-V}$. The best-fitting values of $\alpha$ and $E_{B-V}$ are $\alpha=0.728\pm 0.003$ and $E_{B-V}=0.216\pm 0.004$ mag, respectively.
In Figure \ref{lc_ext}, we assume that the extinction at $z$ band could be neglect, then the solid and dotted lines correspond to the model after and before the extinction correction, respectively. We can see that the extinction at the blue band is significant.
This result implies that the ISM in the host galaxy or the circumburst medium of GRB~250101A has a significant effect on the extinction to the observed afterglow. This is consistent with the general picture that long/Type II GRBs are favored to originate from star formation regions.

\subsubsection{Associated properties between optical and X-ray bands}

\begin{deluxetable*}{ccccc}
  \tablecaption{Spectral properties of Swift/XRT data of GRB~250101A at different time slices \label{XRT_table}}
  \tablecolumns{5}
  
  \tablehead{
    \boldmath$\boldsymbol{T_0+T_{\rm start}}$\unboldmath &
    \boldmath$\boldsymbol{T_0+T_{\rm end}}$\unboldmath &
    \boldmath$\boldsymbol{T_0+T_{\rm middle}}$\unboldmath &
    \boldmath$\boldsymbol{\Gamma_{\rm X}}$\unboldmath &
    \textbf {Flux} at (0.3-10) keV  \\ 
   (s) & (s) & (s)& &($10^{-12}  \text{${\rm erg~cm^{-2}~s^{-1}}$}$) 
  }
  \startdata 
  156.93 & 372.46 & 264.70 & $1.6^{+0.5}_{-0.3}$ & $57^{+17}_{-13}$ \\
  372.46 & 2633.59 & 1503.03 & $1.9^{+0.3}_{-0.2}$ & $34^{+5}_{-4}$  \\
  2633.59 & 6256.46 & 4445.03 & $1.8^{+0.3}_{-0.3}$ & $8.9^{+1.5}_{-1.3}$  \\
  6256.46 & 27210.45 & 16733.46 & $2.2^{+0.4}_{-0.4}$ & $3.2^{+0.9}_{-0.6}$  \\
\enddata
\tablecomments{$T_{\rm start}$ is the start time of the time slice, $T_{\rm end}$ is the end time of the time slice. The middle time $T_{\rm middle}$ is defined as $(T_{\text{start}} + T_{\text{end}})/2$. The results of spectral analysis have been corrected for both the Galactic and intrinsic HI absorption \citep{Evans2009}. The photon index $\Gamma_{\rm X}$ and the XRT flux at $0.3 - 10$ keV were obtained from \url{https://www.swift.ac.uk/xrt_spectra/01278305/}.}
\end{deluxetable*}

Next, we will analyze the wideband properties from X-ray to optical.
The Swift/XRT can automatically change its readout mode based on the source intensity. At high count rates, it operates in Windowed Timing (WT) mode, with a time resolution of $1.8 \times 10^{-3}$ seconds. At lower count rates, the Photon Counting (PC) mode is used, providing full spatial information but with a lower time resolution of 2.5 seconds. Here, we primarily use the PC mode to analyze the X-ray afterglow of GRB 250101A due to its relatively weak X-ray brightness. After manually dividing the XRT data for GRB~250101A into four time periods before $T_0+27210$ s, we submit the time range of every slice to the Swift/XRT website \footnote{\url{https://www.swift.ac.uk/xrt_spectra/01278305/}} \citep[e.g.,][]{Evans2007,Evans2009,Evans2010}. It automatically performs light-curve and spectral analysis of the data, providing the photon index ($\Gamma_{\rm X}$) and the flux for each slice. The provided results of spectral analysis have been corrected for both the Galactic and intrinsic HI absorption \citep{Evans2009} and shown in Table \ref{XRT_table}. The spectral index $\beta_{\rm X}$ for the four slices of GRB 250101A is then derived through the relation of $\beta_{\rm X} = \Gamma_{\rm X} - 1$.

Since the extinction at the optical band is significant as pointed out above, we next combine the observation results of Swift/XRT and Mephisto to check whether GRB~250101A is an ``optical-dark burst'', similar to the recent work on GRB 240825A (Rui-Zhi Li et al., 2025, submitted). 
The standard afterglow model predicts that the high-energy emission has a power-law spectrum under slow cooling process (scaling as $\nu^{-p/2}$).
As the cooling frequency shifts downwards and passes through the optical and X-ray range, the spectral shape will evolve. 
It transitions from $\nu^{-(p-1)/2}$ at lower frequencies to $\nu^{-p/2}$ at higher frequencies. 
Consequently, the spectrum of a GRB afterglow can be represented by a broken power-law model,
\begin{align}
F_{\nu} \propto 
\begin{cases} 
\nu^{-\beta_{\rm O}}, & \nu \ll \nu_{\rm break}, \\
\nu_{\rm break}^{\Delta \beta} \cdot \nu^{-\beta_{\rm X}}, & \nu \gg \nu_{\rm break},
\end{cases}
\end{align}
where $\nu_{\rm break}$ the break frequency between the two segments with different power-law index, and $\beta_{\rm O}=(p-1)/2$ and $\beta_{\rm X}=p/2$ are the spectral indices in the optical and X-ray regimes, respectively, and the parameter $\Delta\beta$ is defined as $\Delta\beta\equiv\beta_{\rm X}-\beta_{\rm O}=1/2$. 
In most scenarios under the slow cooling condition, $\nu_{\rm break}$ is associated with the cooling frequency ($\nu_c$), and the optical and X-ray emissions are mainly at $\nu<\nu_{\rm break}$ and $\nu>\nu_{\rm break}$, respectively.

\begin{figure}
\centering
\includegraphics[width = 1\linewidth, trim = 0 0 0 0, clip]{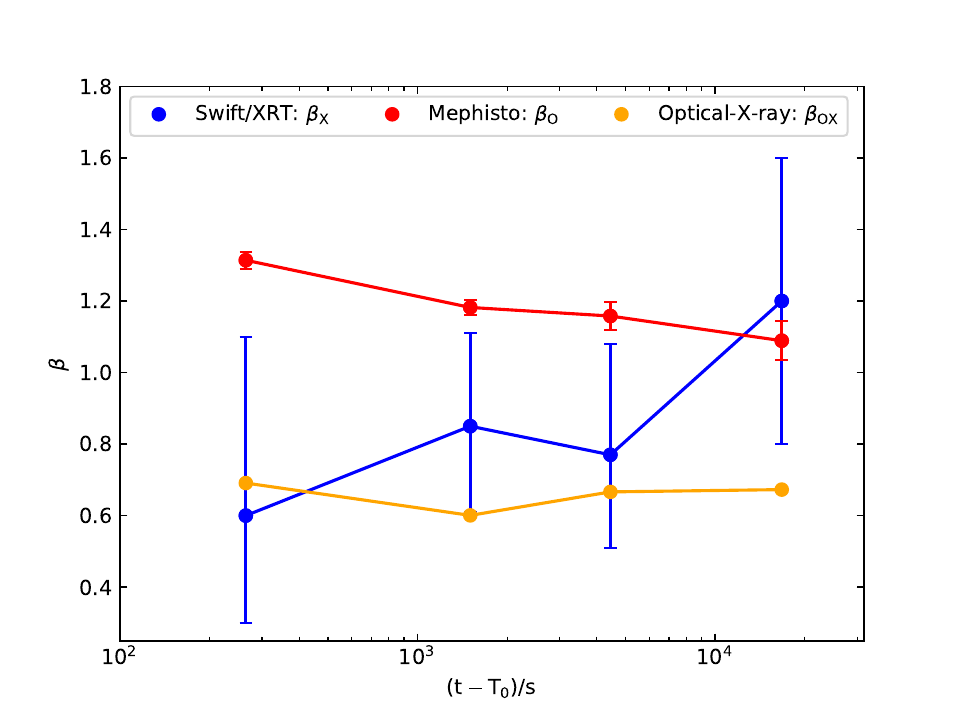}
\caption{The evolution of the spectral indexes of the afterglow at X-ray (blue), optical (red), and optical-to-X-ray (orange) bands, respectively.}\label{beta} 
\end{figure}

Generally, for the typical case with $p\gtrsim2$, the expected spectral index in a wide range between the optical band and X-ray band should be $\beta_{\rm OX}\geqslant0.5$ when $\nu_{\rm break}$ is between the optical and X-ray bands.
Therefore, a GRB can be classified as ``dark'' or ``normal'' based on the optical-to-X-ray spectral index, $\beta_{\rm OX}$, and bursts with $\beta_{\rm OX}<0.5$ are proposed to be optically dark \citep{Jakobsson04}. Another method to identify a ``dark burst'' is based on the criterion of $\beta_{\rm OX}<\beta_{\rm X}-0.5$, which allows for a more subtle diagnostic approach that considers cases where $p < 2$ according to \citet{Horst09}.

The observation of Swift/XRT showed that GRB~250101A has a photon index of $\Gamma_{\rm X}\simeq(1.6-2.2)$ from $\sim T_0+265$ s to $\sim T_0+16733$ s for PC mode, implying $\beta_{\rm X}=\Gamma_{\rm X}-1\sim(0.6-1.2)$ in the same period. Meanwhile, the spectral index $\beta_{\rm X}$ increases over time, as shown in Figure \ref{beta}.
Based on the fluxes at X-ray and optical in the same period, we can calculate the spectral index $\beta_{\rm OX}$ in the wide range. Since the optical band is narrow, we take the flux and the frequency of the $g$ band as the fiducial values. Meanwhile, the flux of X-ray at (0.3-10) keV is shown in Table \ref{XRT_table}. The evolution of $\beta_{\rm OX}$ is shown by the orange line in Figure \ref{beta}. We can see that $\beta_{\rm OX}\gtrsim 0.5$ is satisfied in most periods, suggesting that GRB~25010A is more likely a normal burst based on the criterion of \citet{Jakobsson04}. On the other hand, based on the criterion of \citet{Horst09}, one has $\beta_{\rm X}-\beta_{\rm OX}\lesssim0.5$ in this period, also implying that GRB~250101A is like a ``normal'' burst. Overall, the wideband spectrum from optical to X-ray suggests that GRB~250101A is not an ``optical-dark burst'' and the extinction effect only plays an important role in the optical blue band.

\subsubsection{Implication of optical afterglow evolution}\label{break}

There are two significant features of the optical afterglow of GRB~250101A: 1) the early optical afterglow of GRB 250101A exhibited a rise in brightness at the early phase and showed a peak around 240 seconds post-burst \citep{38777}; 2) the optical light curve, especially at $g$ and $i$ band, appeared a structural change near $\sim T_0+3000$ s, as shown in the top panel of Figure \ref{result} and Figure \ref{residual}.

Although the measurement of Mephisto did not reveal the raising phase of the optical afterglow due to the large exposure of 300 s, the observation of  NUTTelA-TAO / BSTI found that GRB 250101A showed a raising phase and reached the peak brightness $m_{g'}=17.3$ mag $m_{r'}=16.6$ mag at $\sim240$ seconds after the burst \citep{38777}, giving a strong constraint on the onset of the afterglow at the blastwave deceleration radius.

For the spherically symmetric uniform medium with a gas number density of $n$, the observed deceleration time scale can be calculated through \citep[e.g.,][]{Cheng25},
\begin{align}
t_{\rm dec}\simeq186~{\rm s}~(1+z)E_{{\rm iso},52}^{1/3}\Gamma_{0,2}^{-8/3}n_0^{-1/3},
\end{align}
where $E_{\rm iso}$ the isotropic kinetic energy of the afterglow, $\Gamma_0$ the initial Lorentz factor of the ejecta \citep{Zhang18}. Considering that the peak time of optical afterglow corresponds to the deceleration time (i.e., $t_{\rm dec}\sim240~{\rm s}$) and using the redshift of $z\simeq2.481$ \citep{38776}, one can derive the Lorentz factor $\Gamma_0$,
\begin{align}
\Gamma_0\simeq126(1+z)^{3/8}E_{{\rm iso},52}^{1/8}n_0^{-1/8}t_{\rm dec,2}^{-3/8}\sim145E_{{\rm iso},52}^{1/8}n_0^{-1/8}.\label{Gamma0}
\end{align}

Next, combining Figure \ref{result} and Figure \ref{residual}, we can see that the brightness moves downwards near $\sim T_0+3000$ s. This feature is significant at the $i$ band due to its relatively low uncertainty.
However, the temporal slope does not significantly change when we fit the optical light curve with a broken power-law evolution.
As highlighted by \citet{Swenson13}, it is imperative to employ a statistically robust methodology to ensure that an identified break feature is a true signal rather than an artifact of background noise.  To fulfill this requirement, we implemented the publicly available \texttt{R} package \texttt{strucchange} \citep{R_Team,JSSv007i02} and its associated \texttt{breakpoints} function \citep{ZEILEIS2003109}.  The \texttt{breakpoints} function is of particular interest as it utilizes dynamic programming to compute the optimal number of breakpoints in a time series analysis.

\begin{figure}
    \centering
    \includegraphics[width=1\linewidth]{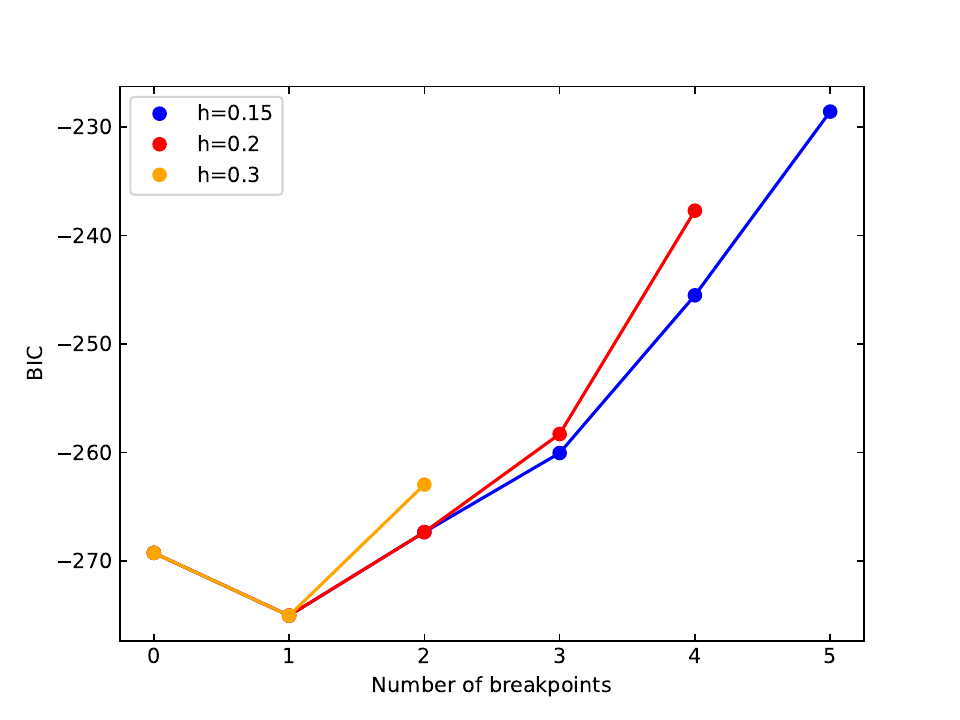}
    \caption{BIC values of different numbers of breakpoints given by the structural analysis for $h$ = 0.15 (blue), 0.2 (red), 0.3 (orange), respectively. This figure shows that the lowest BIC values of these three cases are at the same one breakpoint and the breakpoint time corresponds to $T_0+2924^{+112}_{-190}$ s.}
    \label{fig:bp_BIC}
\end{figure}

In order to obtain the breakpoints based on the measurement at all bands, we first normalized the Mephisto data with the same spectral index of $\beta_{\rm obs}$ and applied the normalized data for structural analysis with \texttt{breakpoints} function of \texttt{R} package. In the \texttt{breakpoints} function, one of the arguments, the minimal segment size $h$, is defined as a fraction of the segment relative to the data size.
A large $h$ can reduce the detection frequency and increase the detection accuracy \citep{JSSv007i02,Bai2003}. To obtain more general results, we set $h=0.15,0.2,0.3$, respectively. The \texttt{breakpoints} function would calculate the Bayesian Information Criterion (BIC) to determine the optimal number of breakpoints and their corresponding positions. We display the BIC results in the Figure \ref{fig:bp_BIC},
and it shows that the lowest BIC values of these three cases are at the one breakpoint and the breakpoint time is 
\begin{align}
    t_{\rm break}\simeq T_0+2924^{+112}_{-190} \ \rm s,
\end{align}
with a confidence error of 90\%.
This breakpoint time has been shown in the vertical grey line in Figure \ref{residual}. 

As shown in Figure \ref{residual}, the structural change at $t_{\rm break}$ is significant at the $g$ band and the $r$ band.
At $uvrz$ bands, the structural change is not easily detectable for a single band due to their relatively large systematic uncertainties. Furthermore, we find that there is no significant color evolution (with a difference of the spectral index of $|\Delta\beta|\lesssim0.11$) before and after the breakpoint $t_{\rm break}$, suggesting that it originate from the kinematic changes rather than spectral evolution.

In physics, an achromatic structural change in the light curve of an afterglow could be due to the energy injection \citep{Dai98} or the jet break \citep{Dermer00,Huang02,Zhang09}. In both scenarios, the temporal slopes before and after the breakpoint $t_{\rm break}$ should be significantly changed. 
However, for GRB~250101A, the slope change before and after $t_{\rm break}$ is very small but the intercept change is significant, as shown in the top panel of Figure \ref{result} and Figure \ref{residual}. Such an interesting feature might be due to a sudden change in the ambient density.
Since the uncertainties at other bands at the late time are large more or less. We mainly focus on the measurement of the Mephisto's $i$ band.
The observed residual magnitude is $\Delta m\simeq0.35$ mag at $i$ band, leading to a flux ratio of 
\begin{align}
\frac{F_{\nu,2}}{F_{\nu,1}}=10^{-0.4|\Delta m|}\simeq0.72,
\end{align}
where $F_{\nu,1}$ and $F_{\nu,2}$ the flux density before and after the breakpoint $t_{\rm break}$, respectively, after deducting long-term temporal evolution. For the standard afterglow model with a uniform ISM \citep[e.g.,][]{Sari98}, the flux density in the adiabatic evolution stage satisfies $F_{\nu}\propto n^{1/2}$. Thus, we obtain a density drop at $t_{\rm break}$,
\begin{align}
    \frac{n_{2}}{n_{1}}\sim10^{-0.8|\Delta m|}\simeq0.52.
\end{align}
Such a result suggests that there is a density drop of $\sim50\%$ at a radius of 
\begin{align}
    r_{\rm break}&\simeq2\Gamma_0^2ct_{\rm dec}+\left[\frac{17E_{\rm iso}(t_{\rm break}-t_{\rm dec})}{4\pi m_{\rm p}c n}\right]^{1/4}\nonumber\\
    &\simeq0.13~{\rm pc}~E_{{\rm iso},52}^{1/4}n_0^{-1/4}.
\end{align}
where $m_{\rm p}$ the proton mass and $c$ the light speed in vacuum. 
The two items on the right side of the equal sign correspond to the coasting phase and the decelerating phase, respectively.
This value is also consistent with the result of the detailed dynamic evolution of the standard afterglow discussed in Section \ref{standard}.
The special environment should be attributed to the outflow medium from the progenitor of GRB 250101A, implying an intermittent outflow process or a possible interaction between the progenitor outflow and the ISM.

\subsubsection{Bayesian inference for standard afterglow model}\label{standard}

Due to the lack of radio observation for GRB~250101A, the total isotropic kinetic energy $E_{\rm iso}$ of the afterglow cannot be well constrained just based on the current observation. In the following discussion, we set the radiation efficiency of the prompt emission as $\eta$, then the total isotropic kinetic energy contributing to the multiwavelength afterglow is
\begin{align}
    E_{\rm iso}=\eta^{-1} E_{\gamma,{\rm iso}},
\end{align}
where $E_{\gamma,{\rm iso}}\simeq 1.4\times10^{52}~{\rm erg}$ the isotropic prompt gamma-ray emission energy discussed in Section \ref{prompt}. 

\begin{figure*}
\centering
\includegraphics[width = 0.45\linewidth, trim = 0 0 0 0, clip]{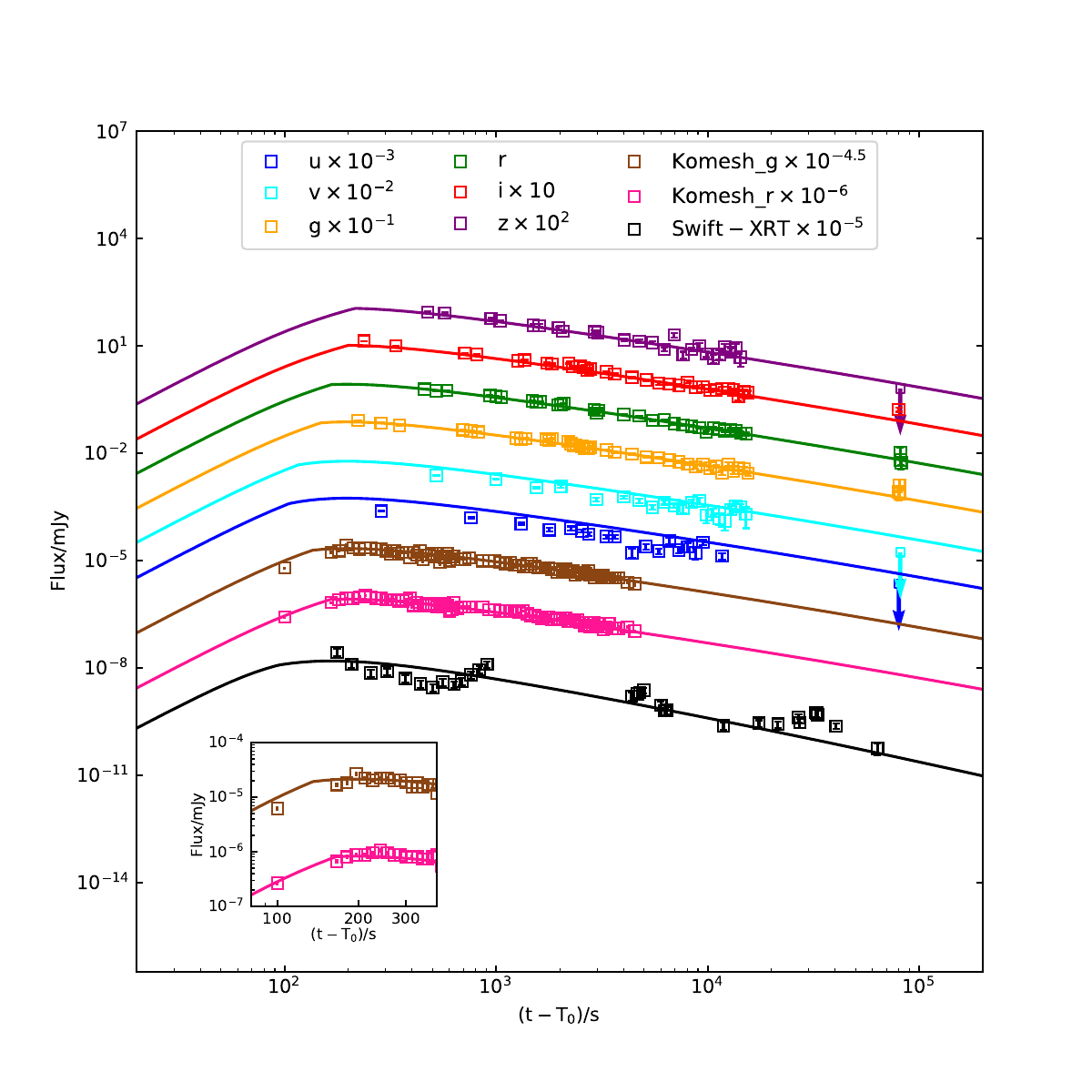}
\includegraphics[width = 0.45\linewidth, trim = 0 0 0 0, clip]{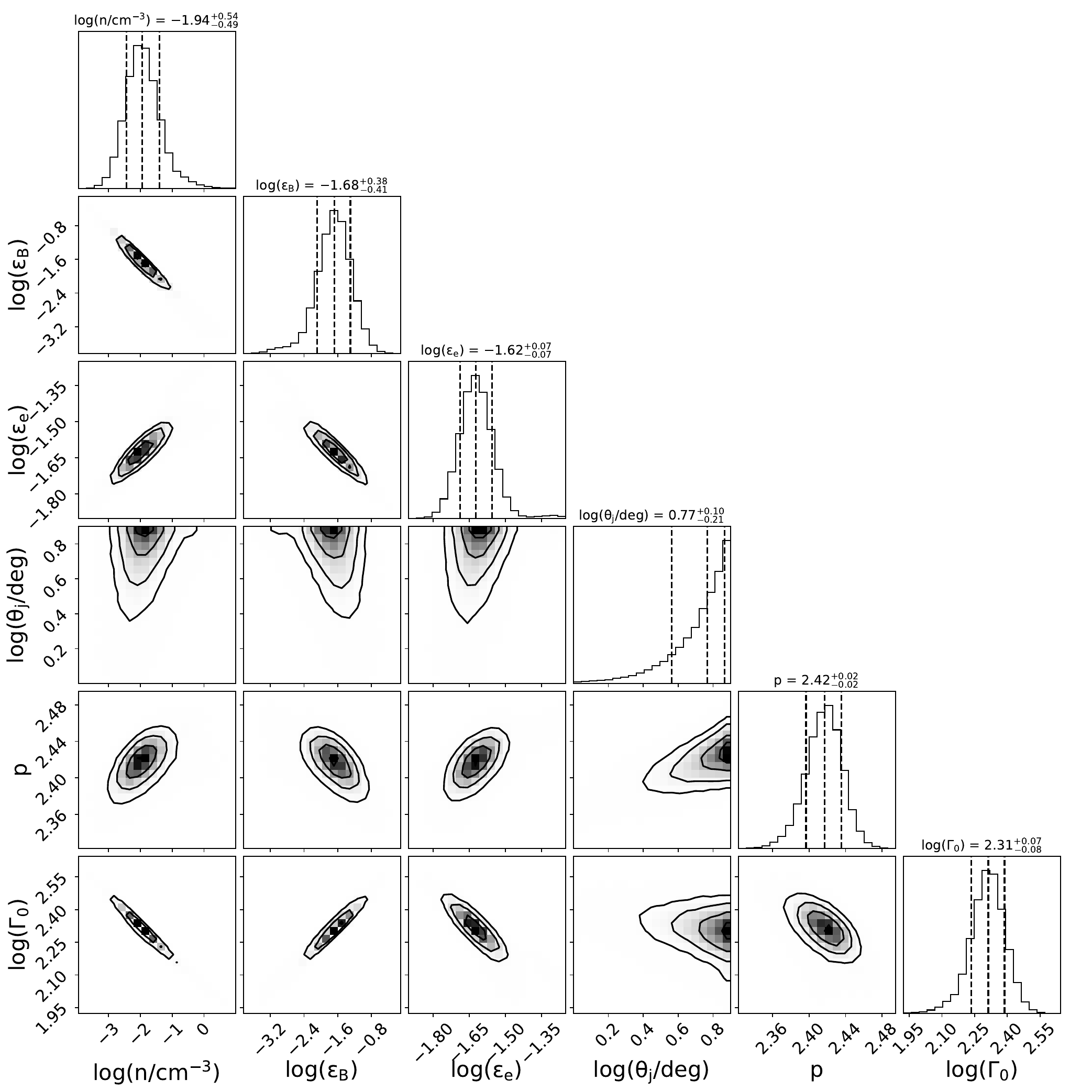}
\includegraphics[width = 0.45\linewidth, trim = 0 0 0 0, clip]{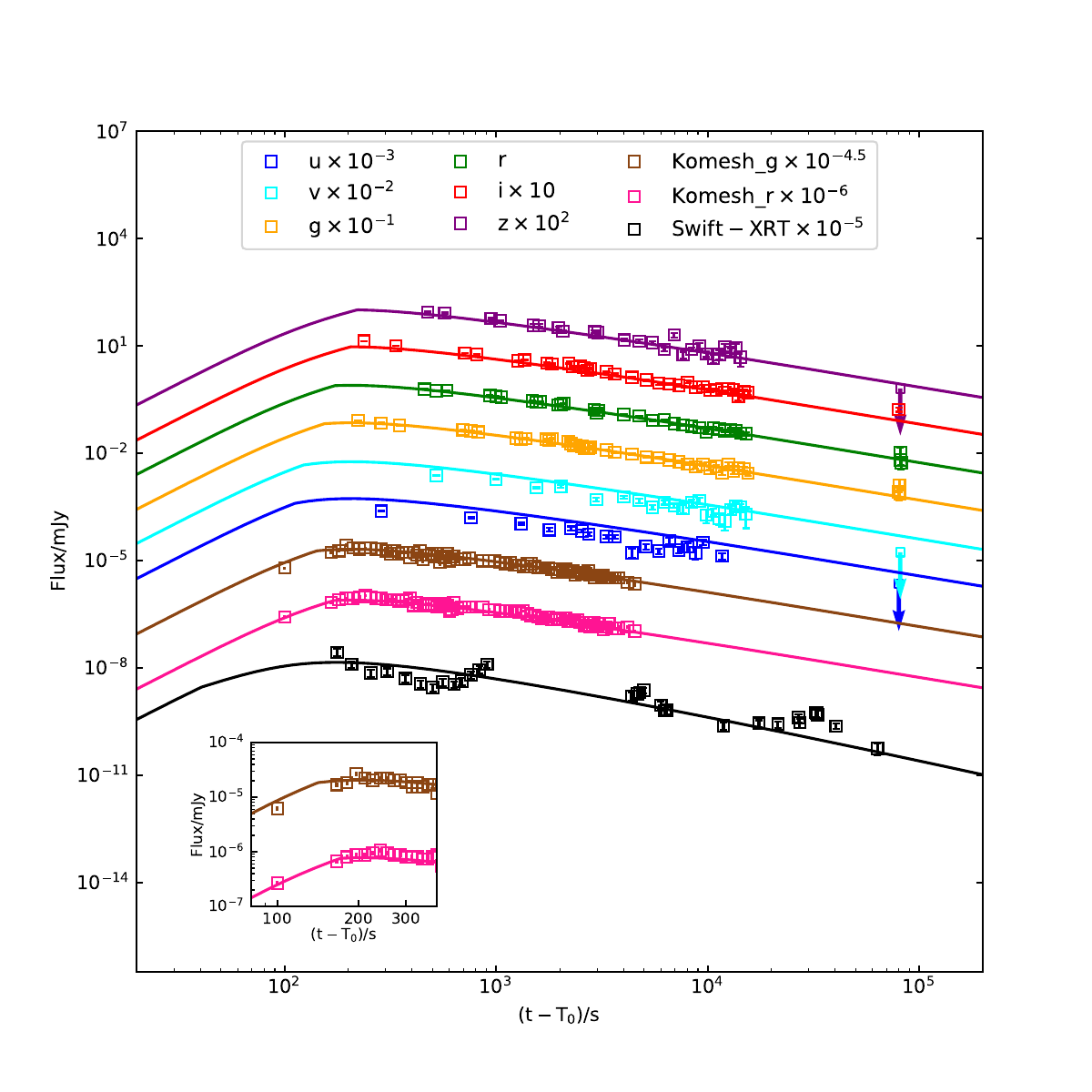}
\includegraphics[width = 0.45\linewidth, trim = 0 0 0 0, clip]{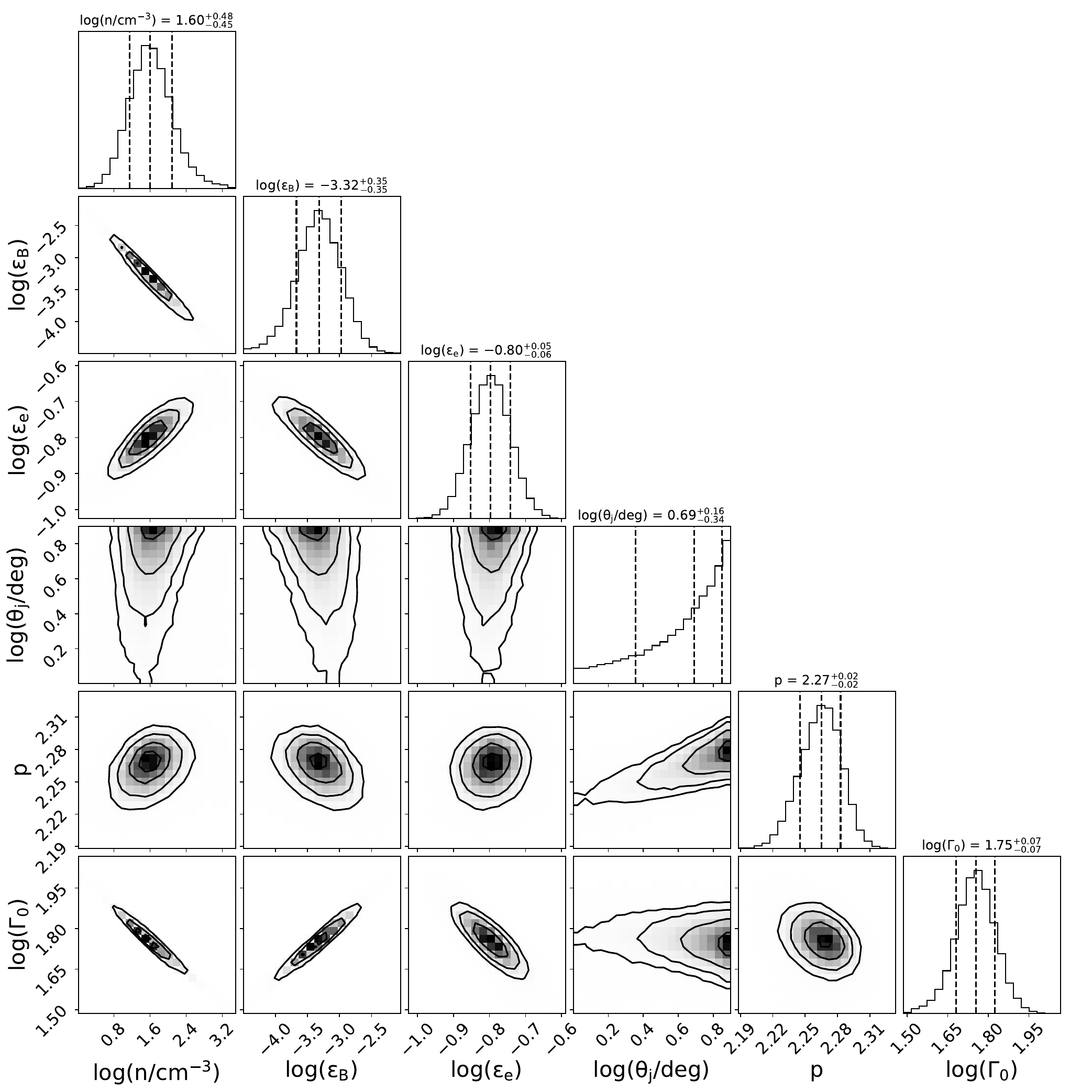}
\caption{Best-fitting results of the standard afterglow model with ISM as the ambient medium. The top and bottom correspond to an efficiency of 5\% and 50\%, respectively. Left panel: Best-fitting results of the measurements of Mephisto, NUTTelA-TAO/BSTI \citep{38777}, and Swift/XRT \citep{Evans2007,Evans2009,Evans2010}. The extinction of $g'r'$ band of NUTTelA-TAO/BSTI are 0.522, and 0.359 mag, respectively.
Right panel: Two-dimensional projections of the posterior probability distributions of the fitting parameters of the standard afterglow model. Contours show the 0.5, 1, 1.5, and 2$\sigma$ significance levels. The Swift/XRT data can be derived from \url{https://www.swift.ac.uk/burst_analyser/01278305/}. }\label{model1}
\end{figure*}

\begin{deluxetable*}{ccccccc}
  \tablecaption{Best-fitting parameters of the standard afterglow model for GRB 250101A \label{params_table}}
  \tablecolumns{7}
  \tablehead{
    \boldmath$\boldsymbol{\eta}$\unboldmath & 
    \boldmath$\boldsymbol{\log(n/{\rm cm^{-3}})}$\unboldmath & 
    \boldmath$\boldsymbol{\log(\epsilon_{\rm B})}$\unboldmath & 
    \boldmath$\boldsymbol{\log(\epsilon_{\rm e})}$\unboldmath & 
    \boldmath$\boldsymbol{\log(\theta_{\rm j}/\mathrm{deg})}$\unboldmath & 
    \boldmath$\boldsymbol{p}$\unboldmath & 
    \boldmath$\boldsymbol{\log(\Gamma_0)}$\unboldmath
  }
  \startdata 
  $5\%$ & $-1.94^{+0.5}_{-0.5}$ & $-1.68^{+0.4}_{-0.4}$ & $-1.62^{+0.07}_{-0.07}$ & $0.77^{+0.1}_{-0.2}$ & $2.42^{+0.02}_{-0.02}$ & $2.31^{+0.07}_{-0.08}$  \\
  $10\%$ & $-0.87^{+0.5}_{-0.5}$ & $-2.21^{+0.4}_{-0.4}$ & $-1.38^{+0.06}_{-0.06}$ & $0.77^{+0.1}_{-0.2}$ & $2.38^{+0.02}_{-0.02}$ & $2.15^{+0.07}_{-0.07}$  \\
  $20\%$ & $0.17^{+0.5}_{-0.5}$ & $-2.68^{+0.4}_{-0.3}$ & $-1.14^{+0.06}_{-0.06}$ & $0.76^{+0.1}_{-0.3}$ & $2.34^{+0.02}_{-0.02}$ & $1.98^{+0.07}_{-0.07}$  \\
  $30\%$ & $0.80^{+0.5}_{-0.4}$ & $-2.97^{+0.3}_{-0.4}$ & $-0.99^{+0.06}_{-0.06}$ & $0.74^{+0.1}_{-0.3}$ & $2.31^{+0.02}_{-0.02}$ & $1.88^{+0.07}_{-0.07}$  \\
  $50\%$ & $1.60^{+0.5}_{-0.5}$ & $-3.32^{+0.4}_{-0.4}$ & $-0.80^{+0.05}_{-0.06}$ & $0.69^{+0.2}_{-0.3}$ & $2.27^{+0.02}_{-0.02}$ & $1.75^{+0.07}_{-0.07}$  \\
\enddata
\tablecomments{$\eta$ is the radiation efficiency of the prompt emission, $n$ is the medium number density of the ambient medium, $\epsilon_{\rm B}$ and $\epsilon_{\rm e}$ are the fraction of shock energy transferring into magnetic field and electrons, respectively, $\theta_{\rm j}$ is the opening angle of the relativistic jet, $p$ is the index of the power-law energy distribution of electrons, and $\Gamma_0$ is the initial Lorentz factor of the GRB jet.}\label{fitting}
\end{deluxetable*}

First, we consider that a relativistic jet with a total kinetic energy of $E_{\rm iso}$ and an opening angle of $\theta_{\rm j}$ propagates in the ISM with a number density of $n$. 
The observed multiwavelength afterglow is produced by the synchrotron emission from the forward shock interacting with the ISM \citep{Sari98}. In this scenario, a set of hydrodynamical equations could describe the dynamical evolution of the relativistic jet of a GRB \citep[e.g.,][]{Huang2000},
\begin{equation}
    \frac{dR}{dt} = \beta \rm c \Gamma(\Gamma+\sqrt{\Gamma^2-1}) ,
    \label{dR/dt}
\end{equation}
\begin{equation}
    \frac{dm}{dR} = 2\pi R^2(1-\cos{\theta_{\rm j}})n m_{\rm p},
    \label{dm/dR}
\end{equation}
\begin{equation}
    \frac{d\theta_{\rm j}}{dt}=\frac{c_{\rm s}(\Gamma+\sqrt{\Gamma^2-1})}{R},
    \label{d0/dt}
\end{equation}
\begin{equation}
    \frac{d\Gamma}{dm} = -\frac{\Gamma^2-1}{M_{\rm ej}+\epsilon m+2(1-\epsilon)\Gamma m},
    \label{dr/dm}
\end{equation}
where $R$ the shock radius, $m$ the swept-up mass, $M_{\rm ej}=E_{\rm iso}(1-\cos{\theta_{\rm j}})/2(\Gamma_0-1)c^2$ the ejecta mass \citep{2023ApJ...948...30Z}, $\beta=\sqrt{\Gamma^2-1}/\Gamma$ the dimensionless velocity of the shock. In the above equation, $c_{\rm s}$ is the sound speed in the the co-moving frame,
\begin{equation}
    c^2_{\rm s} = \hat{\gamma}(\hat{\gamma}-1)(\Gamma-1)\frac{1}{1+\hat{\gamma}(\Gamma-1)}c^2,
\end{equation}
where $\hat{\gamma} \approx (4\Gamma+1)/3\Gamma$ the adiabatic index \citep{Dai1999}. Due to the synchrotron radiation and expanding of ejecta, the electrons will be cooling with a radiative efficiency described by $t'^{-1}_{\rm syn}/(t'^{-1}_{\rm syn}+t'^{-1}_{\rm ex})$ \citep{Dai1999}, where $t'_{\rm syn}=6\pi m_{\rm e}c/\sigma_{\rm T} B'^2 \gamma_{\rm e,min}$ the synchrotron cooling time, $\sigma_{\rm T}$ the Thompson cross section, $B'$ the magnetic strength in co-moving frame, $\gamma_{\rm e,min}$ the minimum Lorentz factor of accelerated electrons, $t'_{\rm ex}=R/\Gamma c$ the expansion time in the co-moving frame \citep{Dai98}.
We define the ratio between the electron's energy and the shock energy as $\epsilon_{\rm e}$, then the radiative efficiency of the total ejecta can be written as \citep{Dai1999}
\begin{equation}
    \epsilon=\epsilon_{\rm e}\frac{t'^{-1}_{\rm syn}}{t'^{-1}_{\rm syn}+t'^{-1}_{\rm ex}}.
\end{equation}
We solve the dynamical evolution based on Eq.(\ref{dR/dt})-Eq. (\ref{dr/dm}), and calculate the synchrotron radiation of the afterglow following \citet{Sari98}. Since the observation did not exhibit a complex structure, we only involve the process of the forward shock based on the ``Occam's Razor'' principle.

We applied this model to fit the observational data that includes the measurements of Mephisto, Swift/XRT, and NUTTelA-TAO/BSTI. All these data have been corrected for the Galactic extinction and HI absorption (i.e., Galactic and intrinsic contributions) \citep{Evans2009}. 
The best-fitting parameters for $\eta=5\%,10\%,20\%,30\%,50\%$ are in the Table. \ref{params_table}. As an illustration, the best-fitting light curves for $\eta=5\%~{\rm and}~50\%$ are shown in the top panel and bottom panel of Figure \ref{model1}, respectively.

\section{Discussions and Conclusions}\label{conclusion}

GRB~250101A was recently triggered by Swift/BAT at 13:22:50 UT on January 1, 2025, and numerous observatories worldwide conducted follow-up observations. In this work, we presented and analyzed the multiwavelength properties of GRB~250101A and its afterglow, based on the observations made with Swift/BAT, Swift/XRT, Fermi/GBM, and Mephisto.
For the prompt gamma-ray emission, we performed the data analysis of Swift/BAT and the Fermi/GBM. 
The duration of GRB~250101A is measured to be $T_{90}=32~{\rm and}~20$ s for Swift/BAT and Fermi/GBM, respectively.
A time-integrated joint spectral analysis showed that the low-energy photon spectral index is $\alpha_{\gamma}=-1.18$ and the peak energy is $E_{\rm p}=33~{\rm keV}$. Based on the averaged gamma-ray flux and the redshift, the isotropic energy of GRB~250101A was estimated to be $E_{\gamma,{\rm iso}}=1.4\times10^{52}~{\rm erg}$. In the Amati relation, GRB~250101A lies well within the $1\sigma$ region of the Type II GRB track, although its rest-frame peak energy is relatively lower than that of the other Type II GRBs. The soft spectrum of the prompt gamma-ray emission of GRB~250101A suggests that it is an X-ray-rich or X-ray-dominated GRB, with intrinsic properties indicating that it is relatively softer than most classical GRBs. 

For optical observation, Mephisto's measurement began 197 seconds after the Swift/BAT trigger, and the light curves in the $uvgriz$ bands roughly follow a single power-law decay in the long term. 
For the optical light curves in $uvgriz$ bands, we find that the decay of the flux density of the early optical afterglow satisfies $F_{\nu,{\rm obs}}\propto t^{-\alpha_{\rm obs}}\nu^{-\beta_{\rm obs}}$ with a temporal decay index of $\alpha_{\rm obs}=0.760\pm0.003$ and a spectral index of $\beta_{\rm obs}=1.207\pm0.014$. 
Assuming that the measured temporal decay index is intrinsic, $\alpha=\alpha_{\rm obs}=0.760$, and the afterglow propagates in the uniform ISM, for slow cooling, an electron spectral index of $p=2.0$ is required and the intrinsic spectral index of the afterglow directly inferred from $\alpha_{\rm obs}$ is $\beta(\alpha_{\rm obs})=0.51$. Thus, our measured spectral index of $\beta_{\rm obs}=1.207$ is significantly larger than those directly inferred from $\alpha_{\rm obs}$ in the standard afterglow model. This result suggests that the blue parts of the optical afterglow of GRB~250101A should be partially absorbed by the circumburst medium, leading to a color excess of $E_{B-V}=0.216$ mag. Such a feature is similar to the case of GRB~240825A recently reported by \citet{Cheng25}. 
The significant extinction effect of GRB~250101A at the blue band might imply the predominance of small grains in the local environments near the GRB source \citep[e.g.,][]{Heintz17,Zafar18}. Due to the rotational disruption and alignment of dust grains by radiative torques induced by afterglows, optical-near-infrared extinction decreases and ultraviolet extinction increases \citep{Hoang20}.

Furthermore, we measured the spectral indexes of the X-ray afterglow and the optical afterglow from $T_0+265$ s to $T_0+16733$ s. The spectral index at the X-ray band is $\beta_{\rm X}\simeq(0.6-1.2)$, and it increases over time. The spectral index in a wide range between the optical band and X-ray band is $\beta_{\rm OX}\gtrsim0.5$, meanwhile, one has $\beta_{\rm X}-\beta_{\rm OX}\lesssim0.5$. This result suggested that GRB~250101A is not likely an ``optical-dark burst'', and the absorption effect pointed out above only plays an important role in the optical blue band.

For GRB~250101A, two main features of its optical afterglow were found: 1) the light curve rose at the early phase and showed a peak around 240 seconds post-burst \citep{38777}; 2) there is a structural change near $T_0+2924$ s. The former feature gave a strong constraint on the ejecta Lorentz factor of $\Gamma_0\sim145E_{{\rm iso},52}^{1/8}n^{-1/8}$. For the latter feature, the BIC exhibits that there is one breakpoint in the light curve and the corresponding breakpoint time is $t_{\rm break}=T_0+2924$ s. This suggested that there is a significant density drop of the ISM, $n_2/n_1\sim0.52$ at $r_{\rm  break}\simeq0.13~{\rm pc}~E_{{\rm iso},52}^{1/4}n_0^{-1/4}$.
At last, we used the standard forward shock interacting
with the ISM to fit the measurement of GRB~250101A. Due to the lack of radio observation,
the total isotropic kinetic energy of the afterglow
cannot be well constrained. Different cases have been discussed and the best-fitting results are shown in Table \ref{fitting}.

In conclusion, GRB 250101A exhibits some special signs in both prompt emission and afterglow. Its relatively softer prompt emission implies a different electron distribution in the emission region compared to most other GRBs, which might originates from a weaker interaction of the internal shock or a weaker magnetic field in the emission region. From the perspective of afterglow, the density drop of 50\% at $\sim0.13~{\rm pc}$ might be due to an intermittent outflow of the progenitor or a possible interaction between the outflow and the ISM. These signs indicate that the progenitor of GRB 2501101A has some peculiarities in the GRB population.

\section*{Acknowledges}

Mephisto is developed at and operated by the South-Western Institute for Astronomy Research of Yunnan University (SWIFAR-YNU), funded by the ``Yunnan University Development Plan for World-Class University'' and ``Yunnan University Development Plan for World-Class Astronomy Discipline''. The authors acknowledge support from the ``Science \& Technology Champion Project'' (202005AB160002) and from two ``Team Projects'' -- the ``Innovation Team'' (202105AE160021) and the ``Top Team'' (202305AT350002), all funded by the ``Yunnan Revitalization Talent Support Program''.
This work is also supported by the National Key Research and Development Program of China (2024YFA1611603) and the ``Yunnan Provincial Key Laboratory of Survey Science'' with project No. 202449CE340002.
This work made use of data supplied by the UK Swift Science Data Centre at the University of Leicester.
A part of the numerical computations was conducted on the Yunnan University Astronomy Supercomputer.
Y.P.Y. is supported by the National Natural Science Foundation of China (grant No.12473047), and the National SKA Program of China (2022SKA0130100). 
J.Z. acknowledges financial support from NSFC grant No. 12103063. 
J.Y. is supported by the NSFC grant No. 13001106. 
We thank the anonymous referee for providing helpful comments and suggestions.
We also acknowledge the helpful discussions with He Gao and Jirong Mao.

\appendix

In Table \ref{1.6mdata_table} in this appendix, we show the photometry data of the optical afterglow of GRB~250101A in the $uvgriz$ band of Mephisto, and all measurements are in AB magnitudes and are not corrected for the Galactic extinction. The times correspond to the middle time of each exposure. The uncertainties of the photometric calibration are estimated to be better than 0.03, 0.01, and 0.005 mag for $u$, $v$, and $griz$, respectively.

\begin{longtable}{ccccccc}
  \caption{Photometry data of GRB~250101A in $uvgriz$ band of Mephisto. 
  } 
  \label{1.6mdata_table} \\
  \toprule
  \textbf{MJD} & \textbf{u/mag} & \textbf{v/mag} & \textbf{g/mag} & \textbf{r/mag} & \textbf{i/mag} & \textbf{z/mag} \\
  \midrule
  \endfirsthead
  \multicolumn{7}{c}{{\tablename\ \thetable{} -- continued}} \\
  \toprule
  \textbf{MJD} & \textbf{u/mag} & \textbf{v/mag} & \textbf{g/mag} & \textbf{r/mag} & \textbf{i/mag} & \textbf{z/mag} \\
  \midrule
  \endhead
  \bottomrule
  \endfoot
  \bottomrule
  \endlastfoot
60676.56010 & -- & -- & 17.07${\pm}$0.01 & -- & -- & -- \\
60676.56027 & -- & -- & -- & -- & 16.30${\pm}$0.01 & -- \\
60676.56083 & -- & -- & 17.25${\pm}$0.02 & -- & -- & -- \\
60676.56084 & 18.66${\pm}$0.04 & -- & -- & -- & -- & -- \\
60676.56141 & -- & -- & -- & -- & 16.62${\pm}$0.01 & -- \\
60676.56156 & -- & -- & 17.43${\pm}$0.02 & -- & -- & -- \\
60676.56284 & -- & -- & -- & 17.30${\pm}$0.02 & -- & -- \\
60676.56300 & -- & -- & -- & -- & -- & 16.73${\pm}$0.05 \\
60676.56355 & -- & -- & -- & 17.42${\pm}$0.02 & -- & -- \\
60676.56356 & -- & 18.59${\pm}$0.03 & -- & -- & -- & -- \\
60676.56414 & -- & -- & -- & -- & -- & 16.80${\pm}$0.05 \\
60676.56428 & -- & -- & -- & 17.39${\pm}$0.02 & -- & -- \\
60676.56559 & -- & -- & 17.73${\pm}$0.03 & -- & -- & -- \\
60676.56574 & -- & -- & -- & -- & 17.15${\pm}$0.02 & -- \\
60676.56631 & -- & -- & 17.80${\pm}$0.03 & -- & -- & -- \\
60676.56633 & 19.12${\pm}$0.06 & -- & -- & -- & -- & -- \\
60676.56689 & -- & -- & -- & -- & 17.23${\pm}$0.02 & -- \\
60676.56704 & -- & -- & 17.89${\pm}$0.03 & -- & -- & -- \\
60676.56832 & -- & -- & -- & 17.74${\pm}$0.03 & -- & -- \\
60676.56848 & -- & -- & -- & -- & -- & 17.17${\pm}$0.05 \\
60676.56905 & -- & -- & -- & 17.79${\pm}$0.03 & -- & -- \\
60676.56906 & -- & 18.87${\pm}$0.04 & -- & -- & -- & -- \\
60676.56963 & -- & -- & -- & -- & -- & 17.35${\pm}$0.05 \\
60676.56977 & -- & -- & -- & 17.84${\pm}$0.03 & -- & -- \\
60676.57198 & -- & -- & 18.31${\pm}$0.04 & -- & -- & -- \\
60676.57215 & -- & -- & -- & -- & 17.68${\pm}$0.03 & -- \\
60676.57271 & -- & -- & 18.37${\pm}$0.04 & -- & -- & -- \\
60676.57272 & 19.52${\pm}$0.07 & -- & -- & -- & -- & -- \\
60676.57328 & -- & -- & -- & -- & 17.63${\pm}$0.03 & -- \\
60676.57343 & -- & -- & 18.35${\pm}$0.04 & -- & -- & -- \\
60676.57471 & -- & -- & -- & 18.10${\pm}$0.04 & -- & -- \\
60676.57486 & -- & -- & -- & -- & -- & 17.65${\pm}$0.08 \\
60676.57544 & -- & -- & -- & 18.19${\pm}$0.04 & -- & -- \\
60676.57545 & -- & 19.45${\pm}$0.07 & -- & -- & -- & -- \\
60676.57600 & -- & -- & -- & -- & -- & 17.68${\pm}$0.09 \\
60676.57616 & -- & -- & -- & 18.18${\pm}$0.04 & -- & -- \\
60676.57744 & -- & -- & 18.39${\pm}$0.05 & -- & -- & -- \\
60676.57759 & -- & -- & -- & -- & 17.83${\pm}$0.04 & -- \\
60676.57816 & -- & -- & 18.47${\pm}$0.06 & -- & -- & -- \\
60676.57818 & 20.0${\pm}$0.1 & -- & -- & -- & -- & -- \\
60676.57873 & -- & -- & -- & -- & 17.91${\pm}$0.05 & -- \\
60676.57889 & -- & -- & 18.35${\pm}$0.06 & -- & -- & -- \\
60676.58017 & -- & -- & -- & 18.39${\pm}$0.05 & -- & -- \\
60676.58034 & -- & -- & -- & -- & -- & 17.83${\pm}$0.08 \\
60676.58090 & -- & -- & -- & 18.36${\pm}$0.05 & -- & -- \\
60676.58091 & -- & 19.38${\pm}$0.09 & -- & -- & -- & -- \\
60676.58147 & -- & -- & -- & -- & -- & 18.1${\pm}$0.1 \\
60676.58163 & -- & -- & -- & 18.29${\pm}$0.05 & -- & -- \\
60676.58289 & -- & -- & 18.53${\pm}$0.07 & -- & -- & -- \\
60676.58306 & -- & -- & -- & -- & 17.87${\pm}$0.05 & -- \\
60676.58362 & -- & -- & 18.71${\pm}$0.07 & -- & -- & -- \\
60676.58363 & 19.9${\pm}$0.1 & -- & -- & -- & -- & -- \\
60676.58418 & -- & -- & -- & -- & 18.09${\pm}$0.05 & -- \\
60676.58435 & -- & -- & 18.84${\pm}$0.07 & -- & -- & -- \\
60676.58517 & -- & -- & 18.82${\pm}$0.07 & -- & -- & -- \\
60676.58625 & -- & -- & 18.86${\pm}$0.06 & -- & -- & -- \\
60676.58640 & -- & -- & -- & -- & 18.08${\pm}$0.05 & -- \\
60676.58697 & -- & -- & 18.89${\pm}$0.06 & -- & -- & -- \\
60676.58699 & 20.0${\pm}$0.1 & -- & -- & -- & -- & -- \\
60676.58753 & -- & -- & -- & -- & 18.16${\pm}$0.06 & -- \\
60676.58770 & -- & -- & 19.02${\pm}$0.06 & -- & -- & -- \\
60676.58845 & -- & -- & 19.03${\pm}$0.07 & -- & -- & -- \\
60676.58868 & -- & -- & -- & -- & 18.34${\pm}$0.06 & -- \\
60676.58917 & -- & -- & 19.05${\pm}$0.07 & -- & -- & -- \\
60676.58921 & 20.3${\pm}$0.1 & -- & -- & -- & -- & -- \\
60676.58979 & -- & -- & -- & -- & 18.24${\pm}$0.06 & -- \\
60676.58990 & -- & -- & 18.94${\pm}$0.07 & -- & -- & -- \\
60676.59122 & -- & -- & -- & 18.72${\pm}$0.06 & -- & -- \\
60676.59137 & -- & -- & -- & -- & -- & 18.12${\pm}$0.09 \\
60676.59194 & -- & -- & -- & 18.97${\pm}$0.07 & -- & -- \\
60676.59196 & -- & 20.3${\pm}$0.1 & -- & -- & -- & -- \\
60676.59251 & -- & -- & -- & -- & -- & 18.2${\pm}$0.1 \\
60676.59266 & -- & -- & -- & 18.77${\pm}$0.06 & -- & -- \\
60676.59593 & -- & -- & -- & -- & 18.43${\pm}$0.04 & -- \\
60676.59596 & 20.4${\pm}$0.1 & -- & -- & -- & -- & -- \\
60676.59597 & -- & -- & 19.13${\pm}$0.03 & -- & -- & -- \\
60676.59956 & -- & -- & -- & -- & 18.62${\pm}$0.04 & -- \\
60676.59958 & 20.5${\pm}$0.1 & -- & -- & -- & -- & -- \\
60676.59959 & -- & -- & 19.31${\pm}$0.04 & -- & -- & -- \\
60676.60395 & -- & 20.10${\pm}$0.09 & -- & -- & -- & -- \\
60676.60396 & -- & -- & -- & 19.07${\pm}$0.03 & -- & -- \\
60676.60422 & -- & -- & -- & -- & -- & 18.69${\pm}$0.09 \\
60676.60829 & -- & -- & -- & -- & 18.84${\pm}$0.05 & -- \\
60676.60831 & 21.6${\pm}$0.3 & -- & -- & -- & -- & -- \\
60676.60832 & -- & -- & 19.44${\pm}$0.05 & -- & -- & -- \\
60676.61242 & -- & -- & -- & -- & -- & 18.8${\pm}$0.1 \\
60676.61244 & -- & 20.4${\pm}$0.2 & -- & -- & -- & -- \\
60676.61245 & -- & -- & -- & 19.17${\pm}$0.04 & -- & -- \\
60676.61655 & 21.1${\pm}$0.2 & -- & -- & -- & -- & -- \\
60676.61657 & -- & -- & 19.63${\pm}$0.06 & -- & -- & -- \\
60676.61684 & -- & -- & -- & -- & 19.03${\pm}$0.06 & -- \\
60676.62091 & -- & -- & -- & -- & -- & 18.9${\pm}$0.1 \\
60676.62095 & -- & 20.8${\pm}$0.1 & -- & -- & -- & -- \\
60676.62096 & -- & -- & -- & 19.48${\pm}$0.05 & -- & -- \\
60676.62547 & -- & -- & -- & -- & 19.25${\pm}$0.09 & -- \\
60676.62550 & 21.4${\pm}$0.2 & -- & -- & -- & -- & -- \\
60676.62552 & -- & -- & 19.68${\pm}$0.06 & -- & -- & -- \\
60676.62961 & -- & -- & -- & -- & -- & 19.3${\pm}$0.1 \\
60676.62963 & -- & 20.5${\pm}$0.1 & -- & -- & -- & -- \\
60676.62964 & -- & -- & -- & 19.45${\pm}$0.06 & -- & -- \\
60676.63372 & -- & -- & -- & -- & 19.31${\pm}$0.08 & -- \\
60676.63374 & 20.7${\pm}$0.2 & -- & -- & -- & -- & -- \\
60676.63376 & -- & -- & 19.85${\pm}$0.07 & -- & -- & -- \\
60676.63784 & -- & -- & -- & -- & -- & 18.3${\pm}$0.1 \\
60676.63786 & -- & 20.7${\pm}$0.1 & -- & -- & -- & -- \\
60676.63788 & -- & -- & -- & 19.71${\pm}$0.06 & -- & -- \\
60676.64196 & -- & -- & -- & -- & 19.40${\pm}$0.08 & -- \\
60676.64199 & 21.4${\pm}$0.2 & -- & -- & -- & -- & -- \\
60676.64200 & -- & -- & 19.96${\pm}$0.08 & -- & -- & -- \\
60676.64609 & -- & -- & -- & -- & -- & 19.7${\pm}$0.2 \\
60676.64611 & -- & 20.9${\pm}$0.3 & -- & -- & -- & -- \\
60676.64613 & -- & -- & -- & 19.80${\pm}$0.08 & -- & -- \\
60676.65056 & -- & -- & -- & -- & 19.17${\pm}$0.09 & -- \\
60676.65059 & 21.2${\pm}$0.3 & -- & -- & -- & -- & -- \\
60676.65060 & -- & -- & 20.13${\pm}$0.08 & -- & -- & -- \\
60676.65468 & -- & -- & -- & -- & -- & 19.3${\pm}$0.1 \\
60676.65470 & -- & 20.5${\pm}$0.2 & -- & -- & -- & -- \\
60676.65471 & -- & -- & -- & 19.87${\pm}$0.07 & -- & -- \\
60676.65883 & 21.6${\pm}$0.3 & -- & -- & -- & -- & -- \\
60676.65884 & -- & -- & 20.3${\pm}$0.1 & -- & -- & -- \\
60676.65911 & -- & -- & -- & -- & 19.5${\pm}$0.1 & -- \\
60676.66316 & -- & -- & -- & -- & -- & 19.1${\pm}$0.2 \\
60676.66319 & -- & 20.4${\pm}$0.2 & -- & -- & -- & -- \\
60676.66321 & -- & -- & -- & 19.99${\pm}$0.07 & -- & -- \\
60676.66729 & -- & -- & -- & -- & 19.5${\pm}$0.1 & -- \\
60676.66733 & 20.8${\pm}$0.2 & -- & -- & -- & -- & -- \\
60676.66734 & -- & -- & 20.16${\pm}$0.09 & -- & -- & -- \\
60676.67145 & -- & -- & -- & -- & -- & 19.7${\pm}$0.2 \\
60676.67147 & -- & 21.4${\pm}$0.4 & -- & -- & -- & -- \\
60676.67148 & -- & -- & -- & 20.30${\pm}$0.09 & -- & -- \\
60676.67627 & -- & -- & -- & -- & 19.8${\pm}$0.1 & -- \\
60676.67631 & -- & -- & 20.4${\pm}$0.1 & -- & -- & -- \\
60676.68041 & -- & -- & -- & -- & -- & 20.0${\pm}$0.3 \\
60676.68044 & -- & 21.1${\pm}$0.2 & -- & -- & -- & -- \\
60676.68045 & -- & -- & -- & 20.01${\pm}$0.07 & -- & -- \\
60676.68454 & -- & -- & -- & -- & 19.7${\pm}$0.1 & -- \\
60676.68458 & -- & -- & 20.3${\pm}$0.1 & -- & -- & -- \\
60676.68866 & -- & -- & -- & -- & -- & 19.7${\pm}$0.3 \\
60676.68869 & -- & 21.6${\pm}$0.3 & -- & -- & -- & -- \\
60676.68870 & -- & -- & -- & 20.09${\pm}$0.09 & -- & -- \\
60676.69279 & -- & -- & -- & -- & 19.6${\pm}$0.1 & -- \\
60676.69284 & -- & -- & 20.7${\pm}$0.1 & -- & -- & -- \\
60676.69691 & -- & -- & -- & -- & -- & 19.2${\pm}$0.2 \\
60676.69694 & -- & 21.8${\pm}$0.4 & -- & -- & -- & -- \\
60676.69696 & -- & -- & -- & 20.2${\pm}$0.1 & -- & -- \\
60676.70225 & -- & -- & -- & -- & 19.6${\pm}$0.1 & -- \\
60676.70228 & -- & -- & 20.1${\pm}$0.1 & -- & -- & -- \\
60676.70637 & -- & -- & -- & -- & -- & 19.5${\pm}$0.2 \\
60676.70639 & -- & 21.0${\pm}$0.3 & -- & -- & -- & -- \\
60676.70640 & -- & -- & -- & 20.1${\pm}$0.1 & -- & -- \\
60676.71050 & -- & -- & -- & -- & 19.7${\pm}$0.1 & -- \\
60676.71054 & -- & -- & 20.6${\pm}$0.1 & -- & -- & -- \\
60676.71464 & -- & -- & -- & -- & -- & 19.3${\pm}$0.3 \\
60676.71468 & -- & 20.8${\pm}$0.3 & -- & -- & -- & -- \\
60676.71469 & -- & -- & -- & 20.2${\pm}$0.1 & -- & -- \\
60676.71500 & 21.8${\pm}$0.2 & -- & -- & -- & -- & -- \\
60676.71877 & -- & -- & -- & -- & 20.2${\pm}$0.2 & -- \\
60676.71881 & -- & -- & 20.4${\pm}$0.1 & -- & -- & -- \\
60676.72292 & -- & 20.8${\pm}$0.2 & -- & -- & -- & -- \\
60676.72294 & -- & -- & -- & 20.4${\pm}$0.1 & -- & -- \\
60676.72794 & -- & -- & -- & -- & -- & 19.9${\pm}$0.4 \\
60676.72885 & -- & -- & -- & -- & 19.9${\pm}$0.2 & -- \\
60676.72890 & -- & -- & 20.4${\pm}$0.1 & -- & -- & -- \\
60676.73303 & -- & 21.3${\pm}$0.5 & -- & -- & -- & -- \\
60676.73304 & -- & -- & -- & 20.4${\pm}$0.1 & -- & -- \\
60676.73715 & -- & -- & -- & -- & 19.9${\pm}$0.1 & -- \\
60676.73719 & -- & -- & 20.8${\pm}$0.2 & -- & -- & -- \\
60677.48338 & -- & -- & 22.1${\pm}$0.4 & -- & -- & -- \\
60677.48804 & -- & -- & 21.6${\pm}$0.3 & -- & -- & -- \\
60677.48872 & -- & -- & -- & -- & 21.1${\pm}$0.1 & -- \\
60677.48875 & $>23.05$ & -- & -- & -- & -- & -- \\
60677.49416 & -- & -- & 22.2${\pm}$0.3 & -- & -- & -- \\
60677.49828 & -- & -- & -- & 21.7${\pm}$0.3 & -- & -- \\
60677.50225 & -- & -- & -- & 22.3${\pm}$0.3 & -- & -- \\
60677.50242 & -- & -- & -- & -- & -- & ${>21.9}$ \\
60677.50247 & -- & ${>23.3}$ & -- & -- & -- & -- \\
60677.50668 & -- & -- & -- & 22.4${\pm}$0.3 & -- & -- \\

\hline
\end{longtable}


\end{document}